\documentclass[
  journal=largetwo,
  manuscript=article-type,
  year=2023,
  volume=37,
]{cup-journal}

\usepackage{amsmath}
\usepackage[nopatch]{microtype}
\usepackage{booktabs}
\usepackage{amssymb}
\usepackage{hyperref}

\title{Deblending overlapping galaxies in \emph{DECaLS} using Transformer-Based algorithm: a method combining multiple bands and data types}

\author{Ran Zhang}
\affiliation{School of Mechanical, Electrical and Information Engineering, Shandong University
180 Wenhua Xilu, Weihai
264209, Shandong, China }
\email[Ran Zhang]{202237594@mail.sdu.edu.cn}
\author{Meng Liu}
\affiliation{School of Mechanical, Electrical and Information Engineering, Shandong University
180 Wenhua Xilu, Weihai
264209, Shandong, China }
\email[Meng Liu]{liumeng@sdu.edu.cn}
\author{Zhenping Yi}
\affiliation{School of Mechanical, Electrical and Information Engineering, Shandong University
180 Wenhua Xilu, Weihai
264209, Shandong, China }

\author{Hao Yuan}
\affiliation{School of Mechanical, Electrical and Information Engineering, Shandong University
180 Wenhua Xilu, Weihai
264209, Shandong, China }

\author{Zechao Yang}
\affiliation{School of Mechanical, Electrical and Information Engineering, Shandong University
180 Wenhua Xilu, Weihai
264209, Shandong, China }

\author{Yude Bu}
\affiliation{School of Mathematics and Statistics, Shandong University
180 Wenhua Xilu, Weihai, 264209, Shandong, China}

\author{Xiaoming Kong}
\affiliation{School of Mechanical, Electrical and Information Engineering, Shandong University
180 Wenhua Xilu, Weihai
264209, Shandong, China }

\author{Chenglin Jia}
\affiliation{School of Mechanical, Electrical and Information Engineering, Shandong University
180 Wenhua Xilu, Weihai
264209, Shandong, China }

\author{Yuchen Bi}
\affiliation{School of Mechanical, Electrical and Information Engineering, Shandong University
180 Wenhua Xilu, Weihai
264209, Shandong, China }

\author{Yusheng Zhang}
\affiliation{School of Mechanical, Electrical and Information Engineering, Shandong University
180 Wenhua Xilu, Weihai
264209, Shandong, China }

\author{Nan Li}
\affiliation{National Astronomical Observatories, Chinese Academy of Sciences, A20 Datun Road, Beijing 100012, China}

\keywords{methods: data analysis – techniques: image processing - cosmology: observations – galaxies: general} 

\begin{document}

\begin{abstract}
In large-scale galaxy surveys, particularly deep ground-based photometric studies, galaxy blending was inevitable. Such blending posed a potential primary systematic uncertainty for upcoming surveys. Current deblenders predominantly depended on analytical modeling of galaxy profiles, facing limitations due to inflexible and imprecise models. We presented a novel approach, using a U-net structured Transformer-based network for deblending astronomical images, which we term the \emph{CAT-deblender}. It was trained using both RGB and the \emph{grz}-band images, spanning two distinct data formats present in the Dark Energy Camera Legacy Survey (\emph{DECaLS}) database, including galaxies with diverse morphologies in the training dataset. Our method necessitated only the approximate central coordinates of each target galaxy, sourced from galaxy detection, bypassing assumptions on neighboring source counts. Post-deblending, our RGB images retained a high signal-to-noise peak, consistently showing superior structural similarity against ground truth. For multi-band images, the ellipticity of central galaxies and median reconstruction error for \emph{r}-band consistently lie within ±0.025 to ±0.25, revealing minimal pixel residuals. In our comparison of deblending capabilities focused on flux recovery, our model showed a mere 1\% error in magnitude recovery for quadruply blended galaxies, significantly outperforming SExtractor's higher error rate of 4.8\%. Furthermore, by cross-matching with the publicly accessible overlapping galaxy catalogs from the \emph{DECaLS} database, we successfully deblended 433 overlapping galaxies. Moreover, we've demonstrated effective deblending of 63,733 blended galaxy images, randomly chosen from the \emph{DECaLS} database.
\end{abstract}

\noindent 

\section{Introduction} \label{sec:intro}
Due to projection effects and interactions among galaxies, blending inevitably occurs when multiple light sources occupy the same region in a given projection during large galaxy surveys. In the context of weak gravitational lensing analyses, blending effects can impact the measurements of shapes and fluxes of blended galaxies, subsequently affecting cosmic shear measurements, potentially degrading the quality of statistical data or introducing biases, as most measurement algorithms assume isolated objects \citep{samuroff2018dark,huang2018characterization}. The outcomes of these effects propagate and alter the statistics of the physical processes being studied. They are anticipated to become one of the main sources of systematic uncertainties in many future scientific surveys, such as the Legacy Survey of Space and Time (LSST \citet{abell2009lsst}) at the Vera C. Rubin Observatory. Blending presents an inverse problem in astronomy, known as deblending, where the goal is to reconstruct the properties of individual sources from a combined, blended observation. The primary strategies for deblending have two aspects. On one hand, deblending will never be perfect due to the inherent underconstraint of the problem: for any recorded photon, we fundamentally do not know its source \citep{melchior2021challenge}. This makes the problem insurmountable, strictly speaking, thus the focus should shift towards understanding the characteristics of blending and its impact on analysis. On the other hand, the better the deblender, the fewer the bias estimates needed for analysis.

Almost all deblending methods are based on the linear blending model, where multiple sources add their radiation flux along the line of sight. This assumption is reasonable in astronomy because blended stars and galaxies usually do not interact physically. They just appear aligned from our observational perspective, and galaxies are generally moderately opaque. One can further assume knowledge of the spatial light distribution of the sources. 
For instance, stars are point sources, whose apparent shape is primarily determined by the point spread function (PSF) of the telescope, so that mainly their luminosity and position need to be determined \citep{melchior2021challenge}. This simplification makes it possible for deblending to successfully separate blended light sources even in highly crowded star scenes \citep{stetson1987daophot,linde1989high,feder2020multiband}. However, galaxies exhibit complex morphologies, so additional assumptions are needed in the deblending method. Fortunately, the morphology of galaxies is not arbitrary. They can be divided into several main morphological features, such as elliptical galaxies and spiral galaxies, and many galaxies can be described by simple parametric models \citep{sersic1963influence} or their approximations \citep{spergel2010analytical,hogg2013replacing}. For multi-source deblending, iterative methods are usually required to improve stability and speed, where the parametric model applies to only one light source at a time. Then, masks can be used to cover pixels believed to belong to other light sources, or the assumed pixel values can be subtracted based on the early fitting results of other light sources \citep{bertin1996sextractor,jarvis2016science,drlica2018dark}. Both of these methods directly take advantage of the independence between the assumed light sources.

Non-parametric methods offer the ability to solve for multiple light sources simultaneously. In the photometric pipeline of the Sloan Digital Sky Survey (\emph{SDSS}), a fitting method has been adopted to implement the linear blending model, which can formally handle any number of light sources \citep{stoughton2002sloan}. This method can be extended using techniques based on matrix decomposition \citep{paatero1994positive,lee1999learning}. To reduce the number of degrees of freedom, morphological heuristics can be used. For example, galaxies can be made symmetrical under 180° rotation or monotonically decreasing from the center through constrained optimization\citep{melchior2018scarlet}. The latest advancements in machine learning in recent years have enabled the use of non-parametric learning methods to observe commonalities between galaxies, thereby expanding and generalizing heuristic constraints. In this regard, some new methods and frameworks have been proposed. In \citet{reiman2019deblending}, a branched adversarial generative network was designed that can deblend overlapping galaxy images in RGB data. This method utilizes the generator and discriminator of adversarial generative networks to learn the commonalities and differences between them by deblending two overlapping galaxy images. \citet{boucaud2020photometry} developed a framework for measuring the luminosity of blended galaxies and uses standard Convolutional Neural Networks (CNN) and U-Net for segmentation. This method can extract luminosity information from blended galaxy images and perform accurate segmentation. \citet{arcelin2021deblending} introduces an algorithm that uses Variational Autoencoder (VAE) neural networks to deblend galaxies in multi-band image data. This method can learn the latent representation of galaxies in multi-band images, thereby accomplishing the deblending task. \citet{wang2022galaxy} designed a neural network framework that can handle an unknown number of galaxies and has no restrictions on galaxy locations. This method can adapt to galaxies of different quantities and locations and perform deblending. However, most current machine learning-based galaxy deblending methods are still exploratory works based on simulated data. These simulated data often cannot fully cover all types and properties of real blended galaxies and may not consider observational conditions, instrument responses, and noise and systematic errors in the data collection process. Therefore, models trained solely on simulated data struggle to fully learn the variations and diversity in real data and cannot be directly applied to galaxy deblending tasks in real data.

In this paper, we introduce a method for the deblending of overlapping galaxies using a Transformer-based generative neural network model called \emph{CAT-deblender}. The principle of deblending is to learn the mapping between the latent probability distribution and the target galaxy distribution, to model the luminosity distribution of celestial bodies in the target galaxy, thus accomplishing the deblending task. In the deblending task, input images usually have a wide spatial range and global relevance. Transformers can effectively model long-distance dependencies. In contrast, Convolutional Neural Networks (CNN) tend to extract local features. Our method demonstrated excellent galaxy deblending performance in real observational datasets, including RGB images and multi-band images. We successfully deblended several genuinely observed blended galaxies. These results verify the feasibility and effectiveness of our method in dealing with real observational datasets. Simultaneously, our research lays a foundation for further research and opens up new possibilities for future development and applications in the field of galaxy deblending.

The structure of this paper is as follows: Section~\ref{sec:datasets} describes in detail the data selection and dataset creation methods we used in our research. Section~\ref{sec:methods} gives an overview of the model architecture we adopted, the implementation strategies, and the experimental settings for model training. In Section~\ref{sec:results}, we present the experimental results, compare them with the industry-standard deblending method—Source Extractor \citep{bertin1996sextractor}, and demonstrate deblending instances for real blended galaxies. Finally, in Section~\ref{sec:summary and discussion}, we discuss our results and briefly outline how to use the model for deblending overlapping galaxies in large-scale survey datasets.
\section{Data} \label{sec:datasets}
In the domain of galaxy deblending, the model predominantly learns the subtle distribution characteristics and latent representations of galaxy images. These characteristics can be harnessed to generate novel galaxy image samples, further shedding light on the statistical attributes and textural intricacies inherent to galaxies. We postulated that specific directions or dimensions within the latent feature space correspond closely to the morphological traits of galaxies. Thus, when constructing a dataset to train the deblending model, it was imperative to ensure comprehensive representation of galaxy morphologies. This section elucidated our dataset architecture, encompassing dataset partitioning, stages of data preprocessing, the creation of artificially blended data, and the meticulous preprocessing of the images.
\subsection{Datasets}
\subsubsection{Data Selection} \label{subsec:Data Selection}
The imaging data employed in this research originates from the \emph{GZD-5} morphological catalog \citep{walmsley2022galaxy}, built upon the \emph{grz}-band data of \emph{DECaLS} DR5 \citep{dey2019overview}. We synthesized RGB photometric images from these three bands following the methodology presented in \citet{walmsley2022galaxy}. Throughout this study, all RGB images are derived by combining the red (r), green (g), and blue (b) channels corresponding to the \emph{grz}-bands \citep{lupton2004preparing}, respectively. To enhance the inclusivity of galaxy morphological attributes, this paper expands upon the thresholds established in \citet{li2022automatic}, aiming to capture a broader spectrum of confirmed galaxy samples. By adhering to the specified thresholds, exemplary galaxies from each category within the \emph{GZD-5} morphological catalog were chosen. As shown in Table~\ref{tab:example_table_1}, galaxies with vote scores that correlate with their respective categories fall within the designated threshold range. In total, our selection yielded 15,484 lenticular galaxies, 17,500 barred spiral galaxies, 17,162 spiral galaxies, 11,075 completely round smooth galaxies, 23,233 between smooth galaxies, and 15,836 cigar-shaped smooth galaxies from an overall count of 100,280 galaxies.

\begin{table*}
\centering
   \caption{The selection of well-sampled galaxies in \emph{GZD-5} is measured. The determination of thresholds is mainly based on Table 2 in \citet{li2022automatic}. Columns T01$\sim$T7 in the third column represent the classification tasks in the GZD-5 decision tree. The fourth column shows the thresholds for selecting clean samples for pre-classification. For instance, to choose galaxies that can identify relatively clean lens-shaped samples, the conditions are $f_{feature/disc}$ \textgreater 0.40, and $f_{edge-on,yes}$ \textgreater 0.45.}
	\label{tab:example_table_1}
\begin{tabular}{lccccl}
\hline
Class & Galaxy                  & Tasks & Selected responses & Thresholds                            & Sample                 \\
\hline
0     & Lenticular              & T01   & Class 1.2          & $f_{feature/disc}$ \textgreater 0.40      & {15484} \\
      &                         & T02   & Class 2.1          & $f_{fedge-on,yes}$ \textgreater 0.45       &                        \\
1     & Barred spiral           & T01   & Class 1.2          & $f_{feature/disc}$ \textgreater 0.40      & {17500} \\
      &                         & T02   & Class 2.2          & $f_{fedge-on,no}$ \textgreater 0.55        &                        \\
      &                         & T03   & Class 3.1          & $f_{a-bar-feature,yes}$ \textgreater 0.45 &                        \\
      &                         & T04   & Class 4.1          & $f_{spiral,yes}$ \textgreater 0.50        &                        \\
2     & Spiral                  & T01   & Class 1.2          & $f_{feature/disc}$ \textgreater 0.45      & {17162} \\
      &                         & T02   & Class 2.2          & $f_{fedge-on,yes}$ \textgreater 0.55        &                        \\
      &                         & T03   & Class 3.2          & $f_{a-bar-feature,no}$ \textgreater 0.55  &                        \\
      &                         & T04   & Class 4.1          & $f_{spiral,yes}$ \textgreater 0.5         &                        \\
3     & Completely round smooth & T01   & Class 1.1          & $f_{smooth}$ \textgreater 0.40             & {11075} \\
      &                         & T07   & Class 7.1          & $f_{completely-round}$ \textgreater 0.48   &                        \\
4     & In between smooth       & T01   & Class 1.1          & $f_{smooth}$ \textgreater 0.40             & {23233} \\
      &                         & T07   & Class 7.2          & $f_{in-between}$ \textgreater 0.55         &                        \\
5     & Cigar-shaped smooth     & T01   & Class 1.1          & $f_{smooth}$ \textgreater 0.40             & {15836} \\
      &                         & T07   & Class 7.3          & $f_{cigar-shaped}$ \textgreater 0.45      &        \\
\hline              
\end{tabular}
\end{table*}
\begin{figure}[!b]
	\includegraphics[width=\columnwidth]{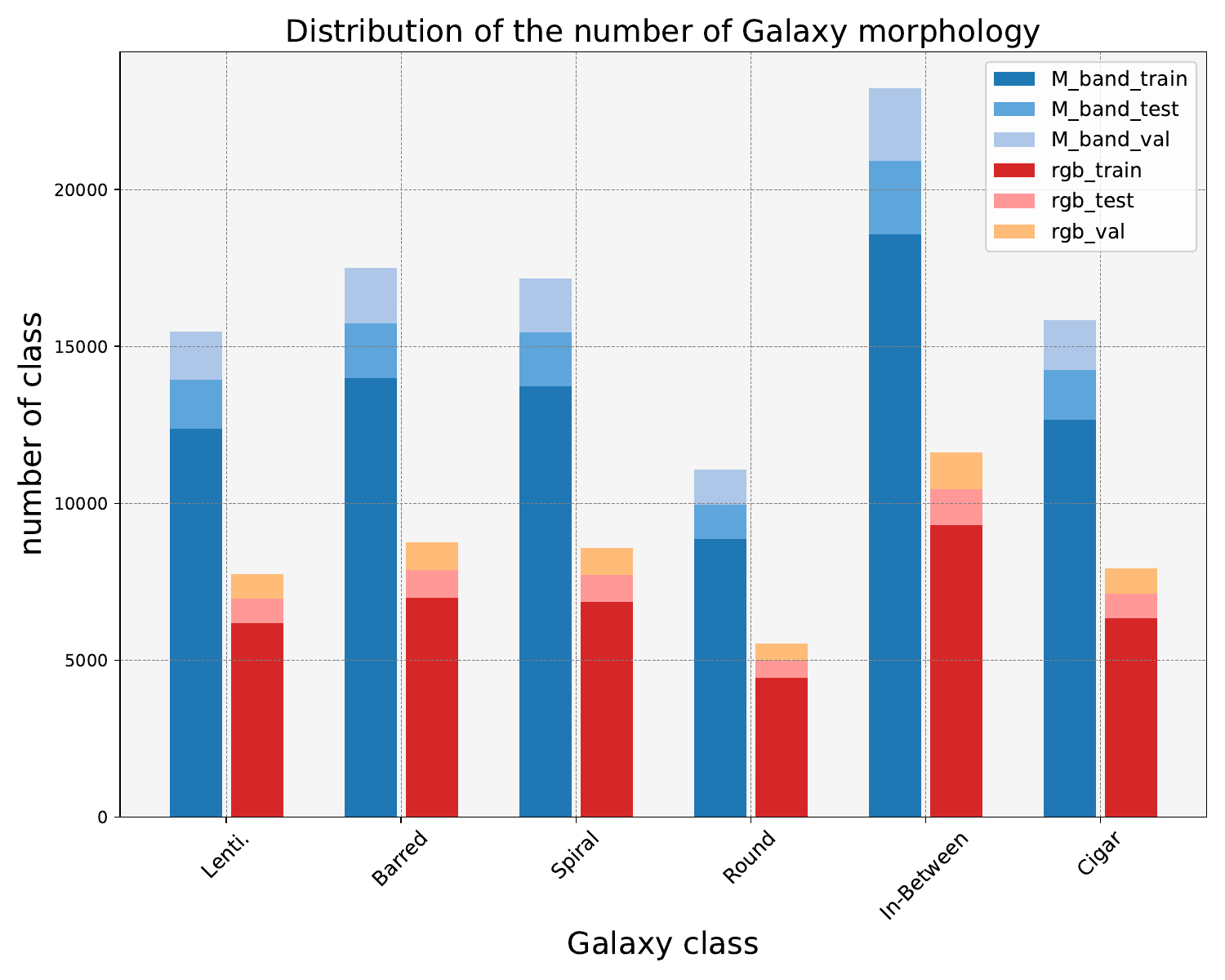}
    \caption{The distribution of the number of galaxies in each class within the dataset, and the count of each galaxy morphology class in two data types: training set, validation set, and test set. The horizontal axis represents the types of galaxies, and the vertical axis represents the quantity of galaxy samples.}
    \label{fig:example_figure}
\end{figure}
The \emph{GZD-5} galaxy catalog facilitates the direct procurement of square FITS cutout images with user-defined pixel dimensions via the \emph{DECaLS} cutout service. This research employs two types of data: RGB data and multi-band data. For consistent display of galaxies in RGB images within a $424\times424$ pixels frame, we calculated the interpolation pixel ratio \emph{s} for FITS cutout images using Equation~(\ref{eq:quadratic_1}).
\begin{equation}
    s=\max(\min(0.04p_{50},0.02p_{90}),0.1)
	\label{eq:quadratic_1}
\end{equation}
Here, $p_{50}$ represents the \emph{Petrosian} $50\%$ light radius, and $p_{90}$ indicates the \emph{Petrosian} $90\%$  light radius. In the case of multi-band data, we procured FITS cutout images with dimensions of $512\times512$, and no scaling modifications were implemented.

A thorough analysis of the dataset revealed a relatively balanced distribution of galaxies across categories. Post data preprocessing, we curated the complete set of galaxy samples. As different training strategies and evaluation criteria are employed for these two types of data during model training, different samples were used for training. For RGB-centric data, considering the imperative of the \emph{Lupton} transformation and hardware optimization, we allocated $50\%$ of the total image samples for training. Conversely, the entirety of multi-band data samples was designated for training. Both datasets were subdivided into training, validation, and test subsets at a proportion of 8:1:1. Figure~\ref{fig:example_figure} depicts the distribution of galaxy classes in the dataset. The training set was used for model training, the validation set for adjusting the model's hyperparameters and monitoring the model, and the test set was used to assess the model's generalization ability on unseen data. It is important to note that the training, validation, and test sets all have the same distribution.
\subsubsection{Data pre-processing}
Given the extreme sensitivity of current large-scale sky survey observations, it's nearly impossible to find an isolated galaxy within a photometric image of appropriate size collected by any survey project. In the field of data-driven machine learning, the quality of images directly affects the effectiveness of the algorithm. Utilizing unprocessed, definitive galaxy images to craft an artificially blended dataset introduces the potential contamination by extraneous entities, be they galaxies or stars, subsequently hampering model training. Consequently, the primary objective in dataset creation is the removal of irrelevant astronomical sources, enabling the algorithm's efficient extraction and assimilation of features.

\begin{figure}
	\includegraphics[width=\columnwidth]{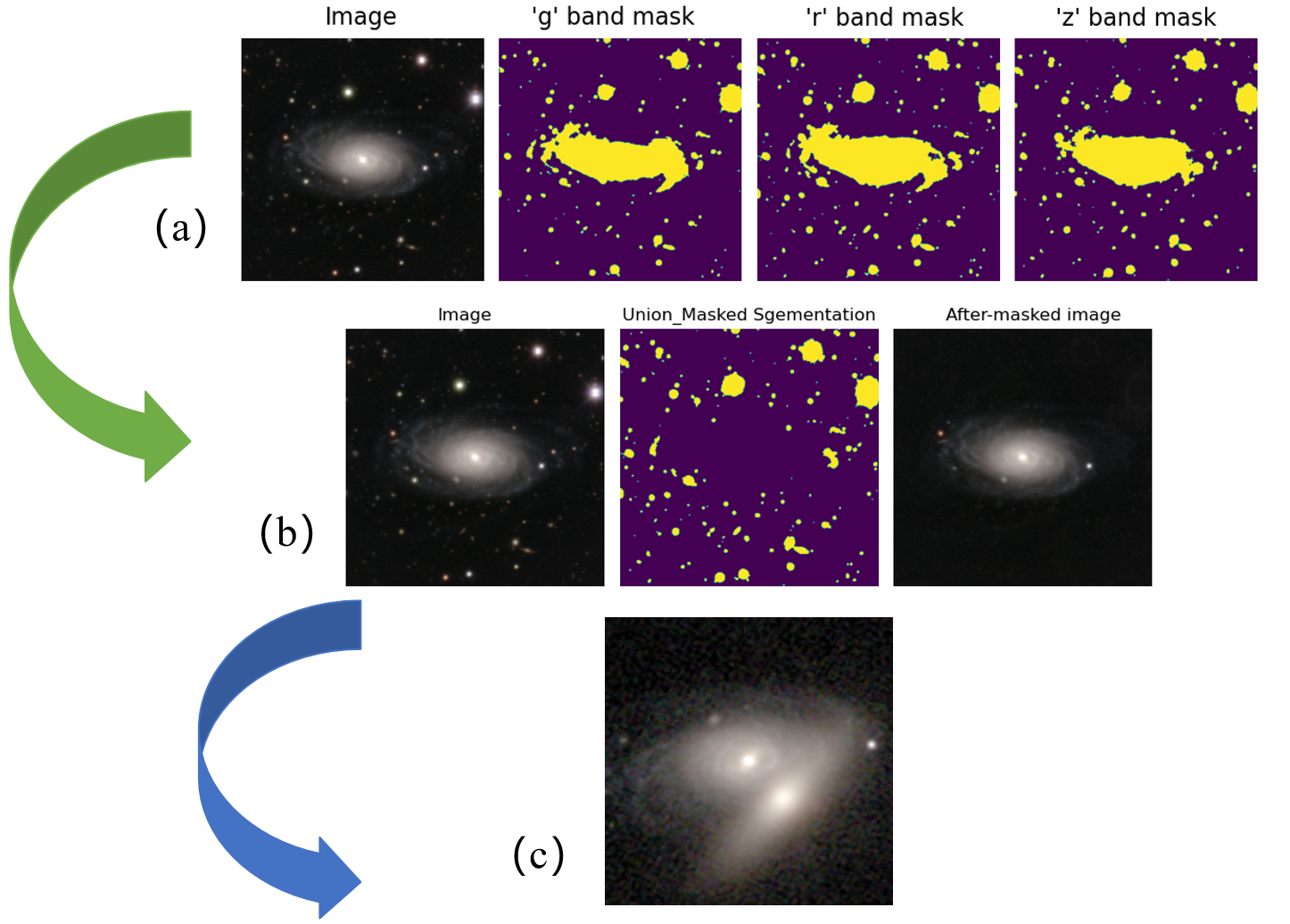}
    \caption{Description of data preprocessing and manual mixing of samples: The (a) process illustrates the creation of masks, where three masks represent the \emph{grz}-bands respectively. The (b) process showcases the masking of irrelevant objects. The mask represents the union of masks for irrelevant objects in the three-band. The third image displays the post-processed image that only contains the target central galaxy. The (c) process demonstrates an example of manual galaxy blending, where the image includes a central target galaxy that needs to be reconstructed by the model and an eccentric galaxy.}
    \label{fig:example_figure_2}
\end{figure}

The data we require centers each isolated galaxy within fixed-sized image blocks. While one might consider overlaying the desired source on a background created with tools like \emph{Photutils} \citep{bradley2016photutils}, differences between this artificial backdrop and the natural background could impact the quality of the resulting images, which may be unfavorable for training Transformer models. To faithfully emulate the background of the original image, we employed the source detection algorithm from \emph{Photutils}, which proved to be an effective tool. This algorithm operates on a foundational principle: it gauges the background intensity of the image, deduces a threshold, and designates a conglomerate of interconnected pixels surpassing this threshold as sources. By setting a threshold of 2$\sigma$ and defining a connection zone of 3 pixels, we created masks for all sources. With these masks, we were able to remove unrelated astronomical entities from the original image, yielding image blocks that solely house the target galaxies and their ambient background.
Figure~\ref{fig:example_figure_2} $(a)$ elucidates the mask generation process, detailing the amalgamation of masks across the \emph{grz}-bands, culminating in a comprehensive mask for all extraneous astronomical sources. Conversely, Figure~\ref{fig:example_figure_2} $(b)$ showcases the masking of these superfluous objects, ultimately yielding images centered on the target galaxies.

Following the procedures outlined, we successfully extracted both isolated target galaxies situated at the center of images and their associated blended galaxies. It is imperative to emphasize that blended galaxies are central to the deblending process. Should they be erroneously classified as isolated galaxies and subsequently included in an artificially blended galaxy dataset, such misclassifications could significantly impact the training of our model and the accuracy of its subsequent inferences. Therefore, to ensure that each image in the finalized dataset exclusively showcases a centrally located isolated target galaxy, we turned to tools like \emph{SExtractor} \citep{bertin1996sextractor} for precise identification and exclusion of overlapping galaxies. We meticulously segregated mixed galaxies from the primary dataset. Through this rigorous processing sequence, we created a dataset containing only images of isolated target galaxies, which are properly centered and sized. This careful selection guarantees the accuracy and dependability of our model during both the training and inference phases.
\subsubsection{Create Artificially blended Galaxy Samples}
The deblending network's training utilizes synthetic samples of blended galaxies. These samples are made by overlaying images of isolated galaxies. It's important to note that this superimposition process does not consider the opacity of the galaxies. Each image may contain 2 to 4 galaxies. The dataset is meticulously organized so that the number of images with two, three, and four galaxies is equally distributed. Any galaxy, regardless of magnitude, can be centered in the blended image and targeted for network reconstruction. Each image's dimensions are represented as (H, W, C), where H, W, and C denote height, width, and channels respectively. For RGB images, H and W are both 424. For multi-band configurations, H and W are both 512. Regardless of the data type, the channel dimension, C, is always set to 3.

For RGB configurations, images are cropped to their central segments with dimensions (H,W,C) = (256,256,3) and then downsampled using bicubic interpolation to (H,W,C) = (128, 128, 3). The center galaxy images remains unchanged, but off-center galaxy images may be horizontally or vertically inverted. Inversions are determined by a Bernoulli distribution with a probability(\emph{p}) of 0.5. The rotation angle, represented by $\theta$, spans a uniform range between 0 and 2$\pi$. Both horizontal and vertical shifts adhere to a uniform distribution within the interval of $dx, dy \in (10,50]$ pixels, while the scaling proportion, denoted by \emph{r}, follows a log-uniform distribution within the range $[1/e, \sqrt{e}]$. Multi-band images uses a similar method; however, the initial multi-band dataset is unmodified to preserve inherent galaxy sizes. These images are then cropped to the central section, matching dimensions (H,W,C) = (128,128,3). These approaches enable the creation of complex galaxy blends, avoiding cases where edge galaxies completely overlap central ones. Figure~\ref{fig:example_figure_2} (c) displays a synthetic galaxy blend, with a central target galaxy ready for model-based reconstruction and an off-center galaxy.

\subsection{Image preprocessing}
For effective model training, it's essential to adjust both types of image data to the range of [-1,1]. The applied normalization method must also allow conversion back to the original range. RGB images and multi-band images employ distinct normalization techniques. $x_{b}$ represents the preprocessing process for each band in multi-band images, where $\langle max(x_{raw},b)\rangle_{b}$ is the average of the maximum pixel value distribution for band $b$ in the input image (each band being a constant), subsequently normalizing each band's original pixels to the range of [-1,1]. $x_{c}$ denotes the preprocessing process for each channel $c$ in RGB images, first normalizing all original pixel values to the range of [0,1] by dividing them by 255. For both types of images, they are then normalized to [-1,1] through two steps of $\sinh^{-1}$ and $\tanh^{-1}$ operations. This processing is necessary, especially for the original multi-band images, as the dynamic range of astronomical images is large, ensuring that images of bright galaxies do not cause numerical instability during training. 

\begin{equation}
    x_{b} = \tanh\left(\sinh^{-1}\left(\frac{x_{\text{raw},b}}{\langle \max(x_{\text{raw},b})\rangle_{b}}\right)\right)
	\label{eq:quadratic_2}
\end{equation}
\begin{equation}
        x_{c} = \tanh\left(\sinh^{-1}\left(\frac{x_{\text{raw},c}}{255}\right)\right)
	\label{eq:quadratic_3}
\end{equation}
\section{METHOD}\label{sec:methods}
 This section offers a succinct overview of the model's fundamental components, emphasizing the architecture of \emph{CAT-deblender}. Commencing with Section~\ref{sec:sec_3_1}, we elucidated the self-attention mechanism. In Section~\ref{sec:sec_3_2}, we delve into the model's basic components, framework, pertinent loss functions, and training setup.
\subsection{Self-attention}\label{sec:sec_3_1}
The self-attention mechanism enables models to discern relationships within a sequence \citep{vaswani2017attention}. It models long-range data dependencies and is effective in tasks like natural language processing (NLP) and computer vision. In NLP, self-attention is typically used to calculate the weighted sum of each word's hidden representation in a sentence, where the weights are determined by each word's attention level to other words in the sentence. This enables the model to learn long-range dependencies between words, which is highly applicable to tasks such as machine translation and question answering. In computer vision, self-attention is usually used to compute the weighted sum of feature maps for each pixel in an image, with the weights determined by each pixel's attention to other pixels in the image. The objective of galaxy deblending is to extract an image focused on the central galaxy from scenes containing multiple overlapping galaxies. This process involves identifying and highlighting the central galaxy while reducing the influence of other overlapping galaxies. The key is to use the non-overlapping parts of the image to predict the pixel ownership in overlapping regions. Given the complexity of spatial relationships in galaxy images, the model needs to consider both the global layout and local details. The self-attention mechanism aids in this by calculating attention scores between different regions, thereby enabling the model to identify features most relevant to the central galaxy, which enhances accuracy and efficiency in galaxy image deblending. Consequently, the self-attention mechanism plays a crucial role in galaxy deblending tasks, particularly in capturing long-range correlations.

\begin{figure*}[!t]
	\includegraphics[width=\textwidth]{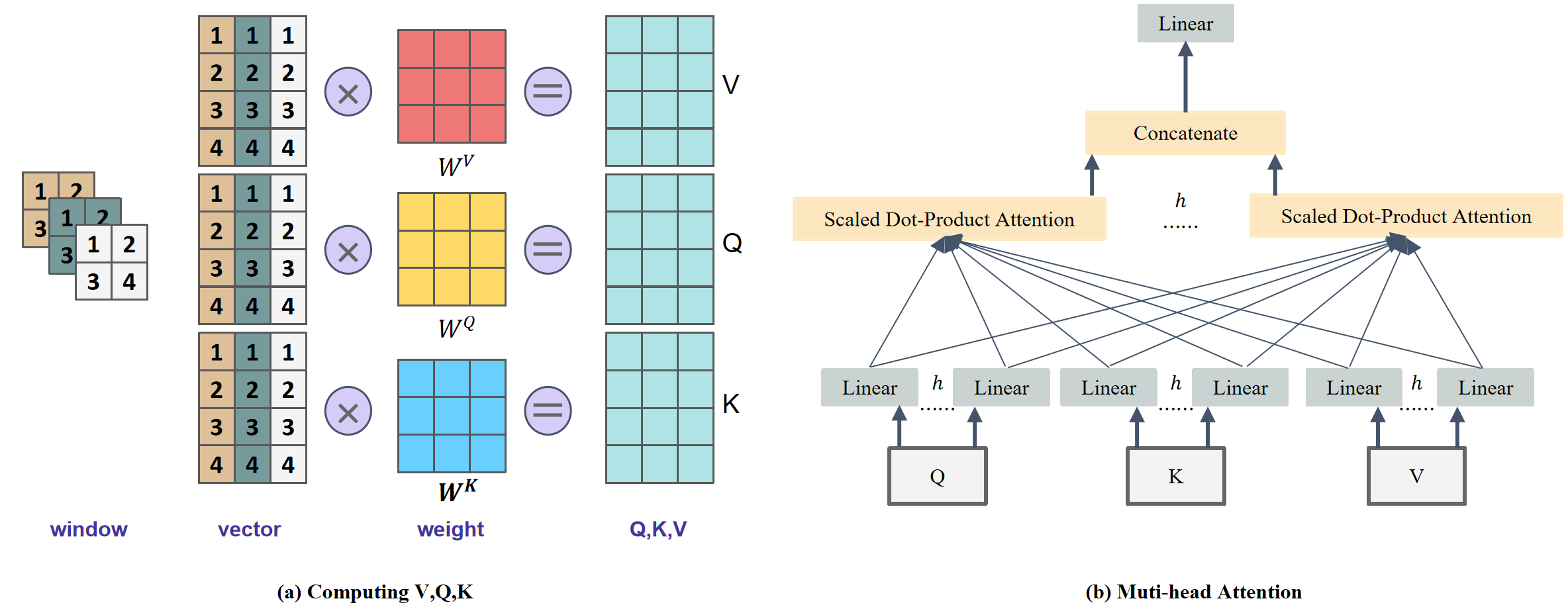}
    \caption{The multi-head attention mechanism as defined in \citet{vaswani2017attention} is described as follows. (a): The input sequence $X$ is multiplied with the weight matrices $W^V$, $W^K$, and $W^Q$, resulting in $V$, $K$, and $Q$ respectively. (b): For multi-head attention, the value $V$, key $K$, and query $Q$ each undergo $h$ different learned linear projections. Subsequently, multi-head attention components generate $h$-dimensional output values in parallel. Finally, they are concatenated and projected, yielding the final value.}
    \label{fig:example_figure_3}
\end{figure*}

The purpose of self-attention is to calculate the similarity among every input in a sequence. It uses three vectors—Query (Q), Key (K), and Value (V)—for this computation. The Query (Q) is used to describe the role of the current position or element in the attention mechanism. The Key (K) represents the importance or relevance of each element to the others. The Value (V) is the actual numerical representation associated with each element. To compute these vectors as three distinct subspaces, the input sequence is projected using three matrices, $W_{Q}$, $W_{K}$, and $W_{V}$. These are weight matrices initialized with independent random values, which the model learns. For instance, given an image of size $M^2$
 pixels, the spatial dimensions (like height and width) of the corresponding tensor are flattened into one dimension, forming a one-dimensional input sequence of length $M^2$
, defined as X. With the three weight matrices $W_{Q}$, $W_{K}$, and $W_{V}$, Q, K, and V are computed as: $Q=XW_{Q}$, $K=XW_{K}$, $V=XW_{V}$. The first step in this process is to multiply each encoder input sequence X with the three weights ($W_{Q}$, $W_{K}$, and $W_{V}$). As shown in Figure~\ref{fig:example_figure_3} $(a)$, we then perform matrix multiplication with $W_{Q}$, $W_{K}$, and $W_{V}$ and take them as inputs to the attention function. The self-attention matrix A, often referred to as scaled dot-product attention, can be written as: $Attention = softmax(QK^{T}/d_{k})$, where $d_{k}$ is the dimensionality of the queries and keys. The attention layer accepts input in the form of these three parameters. They have a similar structure where each element in the sequence is represented by a vector.

In the Transformer architecture, attention modules often consist of several self-attention heads running in parallel, also referred to as multi-head attention mechanism. As shown in Figure~\ref{fig:example_figure_3} $(b)$, this allows the neural network to simultaneously learn from \text{h} different representation subspaces of \text{Q}, \text{K}, and \text{V}. In multi-head attention, keys, values, and queries are linearly projected \text{h} times, i.e., multiple parallel computations can be done, and these results are then concatenated to obtain the final value. In Equation~(\ref{eq:quadratic_4}), $W^{0}$ are the learned output weights, and Equation~(\ref{eq:quadratic_5}) represents the attention matrix computed by each \text{attention head}. The result of this linear transformation is a multi-head attention matrix of dimension $n \times d$. Using multiple attention heads allows the network to learn from a more enriched final representation.
\begin{equation}
\text{MultiHead}(Q, K, V) = [\text{head}_1, \ldots, \text{head}_h]W^0\\
	\label{eq:quadratic_4}
\end{equation}
\begin{equation}
\text{head}_i = \text{Attention}(QW_i^Q, KW_i^K, VW_i^V)
	\label{eq:quadratic_5}
\end{equation}
\begin{figure*}[!t]
    \centering
	\includegraphics[width=\textwidth]{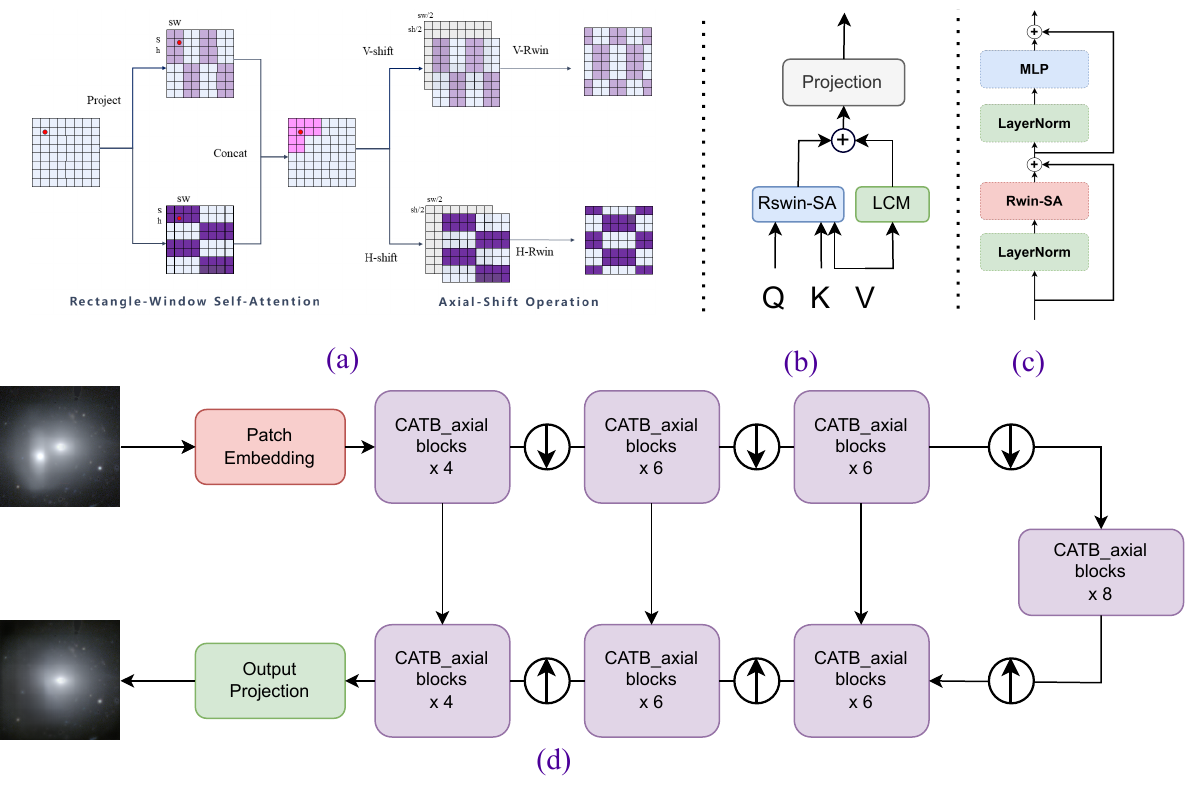}
    \caption{(a): Illustration of the rectangular window self-attention mechanism and axial shift operation. $sh$ and $sw$ are the window height and width of H-Rwin and V-Rwin, respectively. For a pixel (red dot), the attention area is the pink region in the middle feature map. (b): Description of the Local Complementary Module. (c): Explanation of the CATB module. (d): Overview of the structure of \emph{CAT-deblender}.}
    \label{fig:example_figure_4}
\end{figure*}
\subsection{Cross-Aggregation Transformer Model}\label{sec:sec_3_2}
\subsubsection{Cross-Aggregation Transformer Module}
The basic module of the model framework we use is the Cross-Aggregation Transformer Module \citep{chen2022cross}. We introduce the three core designs of this module, including rectangular window self-attention, axial shift operation, and local complementary module.

Rectangular Window Self-Attention(Rwin-SA): We utilize a new window attention mechanism, as shown in Figure~\ref{fig:example_figure_4} $(a)$, called rectangular window self-attention. To alleviate the drawback of excessive time complexity caused by using global self-attention like $ViT$ \citep{dosovitskiy2020image}, this design implements self-attention operation within non-overlapping local rectangular windows, significantly reducing the computational cost. Moreover, the rectangular window is divided into horizontal windows (\text{\(sh > sw\) }, denoted as \text{V-Rwin}) and vertical windows (\text{\(sw > sh\)}, denoted as \text{H-Rwin}), and they are applied in parallel to different attention heads. Through \text{V-Rwin} and \text{H-Rwin}, we can aggregate features across different windows and broaden the attention area without increasing computational complexity, and the rectangular window can capture different features in the horizontal and vertical directions for each pixel. Specifically, given the input $X \in \mathbb{R}^{H*W*C}$, by performing attention operation $h$ times in parallel, for each attention head, we divide $X$ into non-overlapping rectangular windows and compute global self-attention within each rectangular window. For \text{V-Rwin} and \text{H-Rwin}, it's the same, assuming the number of attention heads is even, divide the heads into two parts equally and parallelly execute the first part \text{H-Rwin} and the second part \text{V-Rwin}. Finally, the outputs of the two parts are concatenated along the channel dimension.

Axial Shift Operation: To further expand the receptive field of \text{Rwin-SA} and enable each pixel to gather more information, this module employs axial shift operation of \text{Rwin-SA}. As shown in Figure~\ref{fig:example_figure_4} $(b)$, axial shift includes two shift operations, the horizontal shift referred to as H-Shift, and the vertical shift referred to as V-Shift, corresponding to \text{V-Rwin} and \text{H-Rwin} respectively. The axial shift operation moves the window partition down and to the left by $sh/2$ and $sw/2$ pixels, where $sh/2$ and $sw/2$ are the window height and width of \text{H-Rwin} and \text{V-Rwin}, respectively. In implementation, we cyclically shift the feature map to the lower left direction. Then, we perform \text{V-Rwin} and \text{H-Rwin} on the corresponding shifted feature maps and use a masking mechanism to avoid false interactions between non-adjacent pixels. Then \text{Rwin-SA} concurrently executes \text{V-Rwin} and \text{H-Rwin}, and they are fused together through a projection matrix. Therefore, axial shift can implicitly realize the interaction between horizontal and vertical rectangular windows.

Local Complementarity Module:Transformers are effective at capturing global information and modeling long-distance dependencies between pixels. However, the inductive bias of CNNs (translational invariance and locality) is still indispensable in image restoration tasks. It can aggregate local features and extract basic structures of the image (like corners and edges). To supplement locality for the Transformer, and realize the coupling of global and local, a separate convolution operation, referred to as a local complementarity module, is added in this module during self-attention computation. Unlike previous Transformer approaches, it requires a step to directly convolve on value $V$, as shown in Figure~\ref{fig:example_figure_4} $(b)$. The process can be represented as in Equation~(\ref{eq:quadratic_6}). Here, $Y^1,...,Y^{h/2}$ use the attention head output from H-Rwin, $Y^{h/2},...,Y^{h}$ use the attention head output from V-Rwin, $V \in R^{H*W*C}$ is the direct projection of $X$, and $W^p \in R^{C*C}$ represents the projection matrix for feature fusion.
\begin{equation}
\text{Rswin-SA(X)} = (Concat(\text{Y}^1,\ldots,\text{Y}^h) + Conv(V))\text{W}^p
	\label{eq:quadratic_6}
\end{equation}

Cross-Aggregated Transformer Block:As shown in Figure~\ref{fig:example_figure_4} $(c)$, the \text{CATB} consists of rectangular window self-attention (Rwin-SA) and Multi-Layer Perceptron (MLP), the latter comprising two fully connected layers connected through a \text{GELU} non-linear activation function. Axial shifting operations are conducted on the intervals between two consecutive cross-aggregated Transformer blocks, enhancing the interaction between different windows.
\subsubsection{The overall pipeline of the model}
As depicted in the figure, the overall structure of our proposed \text{CAT-deblender} is a U-shaped hierarchical network \citep{ronneberger2015u}, featuring skip connections between the encoder and the decoder. Specifically, given an image $I_0 \in \mathbb{R}^{3*H*W}$ containing blended galaxies, the \emph{CAT-deblender} initially uses a $3\times3$ convolution layer with \text{LeakyReLU} to extract low-level features $X_0 \in \mathbb{R}^{3*H*W}$. Following the U-shaped structure's design, the feature map goes through three encoder levels. Each stage comprises a stack of CATB Transformer modules and a downsampling layer. The CATB Transformer block leverages a self-attention mechanism and a local complementary module to capture distant dependencies and aggregate local features, reducing computational cost through rectangular window self-attention and axial shift operations on the feature map. In the downsampling layer, we first reshape the flattened features into 2D spatial feature maps, use a $3\times3$ convolution layer to halve the number of channels, and then employ a pixel shuffle layer to double the number of channels and halve the spatial dimensions of the image. Then, a bottleneck stage with a stack of CATB Transformer modules is added at the end of the encoder. In this stage, due to the hierarchical structure, the Transformer block can capture longer dependencies. For feature reconstruction, the decoder also includes three stages. Each stage is composed of an upsampling layer and a stack of CATB Transformer modules similar to the encoder. We use a $3\times3$ convolution layer to double the number of channels, and then a pixel shuffle layer to halve the number of channels and double the image's spatial dimensions. The features input to the CATB Transformer block are the concatenation of the upsampling features and the corresponding features obtained from the encoder via skip connections. The CATB Transformer block is then used to learn image reconstruction. After three decoder stages, we reshape the flattened features into 2D feature maps and apply a $3\times3$ convolution layer to obtain the final output image $I_1 \in \mathbb{R}^{3*H*W}$, which is the target image with the deblended isolated galaxy located at the center.

\subsubsection{Loss Function and Training Settings}
We employed distinct loss functions tailored to specific data types.
For RGB images, the mean squared error (MSE) was computed on a pixel-by-pixel basis, contrasting generator samples $I^{DB}$ with the corresponding ground truth images $I^{PB}$. This approach is analogous to augmenting the effective log-likelihood of pixel intensity data in relation to the network parameters, under the assumption of a Gaussian likelihood function. The pixel-level MSE for an image, defined by its pixel width $W$ and height $H$, is expressed in Equation~(\ref{eq:quadratic_7}).
For multi-band images, the \emph{Charbonnier} \citep{lai2018fast} loss was computed between the generator samples and the ground truth images. This is equivalent to maximizing the effective log-likelihood of the data with respect to the network parameters, given a modified Gaussian likelihood function. For an image with pixel width $W$ and pixel height $H$, the pixel-level Charbonnier error is as shown in Equation~(\ref{eq:quadratic_8}). Here, $e$ remains a constant experimental factor. Our rationale for employing MSE loss with RGB images and Charbonnier loss for multi-band images stems from the distinct nature and requirements of each image type. For RGB images, the emphasis lies on the precision of color representation and overall visual fidelity. The MSE loss, being highly sensitive to minor discrepancies, is adept at discerning nuanced alterations in color and luminance. In contrast, multi-band images, often riddled with noise, outliers, and anomalies, necessitate a more robust loss function. The Charbonnier loss, leveraging the L1 norm for pronounced differences, offers superior robustness, minimizing the influence of aberrations during model training.
\begin{equation}
\text{$l$}_{MSE} = \frac{1}{WH}\sum_{x=1}^{W}\sum_{y=1}^{H}[I_{x,y}^{PB} - G_{\theta G}{(I^{BL}})_{x,y}]^2
	\label{eq:quadratic_7}
\end{equation}
\begin{equation}
\text{$l$}_{Charbonnier} = \frac{1}{WH}\sum_{x=1}^{W}\sum_{y=1}^{H}\sqrt{([I_{x,y}^{PB} - G_{\theta G}{(I^{BL}})_{x,y}]^2+\epsilon^2)}
	\label{eq:quadratic_8}
\end{equation}

To prevent overfitting during the training process of the model, in addition to artificially increasing the amount of training data, we have also adopted techniques such as Droppath \citep{larsson2017fractalnet}, L2 Regularization \citep{krogh1991simple}, and Layer Normalization. Droppath operates by stochastically eliminating specific pathways or branches within the network, reducing both its intricacy and capacity, thus curbing overfitting tendencies. L2 Regularization, often termed as weight decay, prioritizes the selection of the smallest weight vector to address the learning task, effectively dampening extraneous components in the vector. This regularization technique minimizes the influence of static noise on the target, enhancing the model's generalization capacity. Layer Normalization works by normalizing each dimension of a sample's features, constraining the data distribution within a consistent range. This confinement enhances the model's resilience to parameter fluctuations and diminishes sensitivity to parameter choice. We initially trained on RGB data with a batch size of 8 and a set initial learning rate \textit{lr} of $1\times10^{-6}$. This learning rate underwent updates based on the cosine annealing algorithm across 80 epochs. Given that RGB and multi-band data possess congruent dimensions, we employed transfer learning. This involved using the terminal model weights derived from RGB image training to further train the multi-band data deblending model for an additional 30 epochs. The experimental setup utilized an Intel Core i7-12700H processor complemented by an NVIDIA RTX 3090 24G GPU. Implementations were carried out within the PyTorch deep learning framework, utilizing the Python programming language.
\section{Results}\label{sec:results}
\subsection{Reconstruction metrics}
In our investigation into the deblending of overlapping galaxies across two distinct data types, we have employed specific reconstruction metrics tailored to the unique characteristics of each dataset. For RGB images, we applied Peak Signal-to-Noise Ratio (PSNR) and Structural Similarity Index (SSIM) \citep{wang2004image} as metrics to assess the quality of the deblended images extracted from the deblender, comparing them with real images. Conversely, for multi-band images, the evaluation pivoted on the quantification of physical parameters. Specifically, we examined ellipticity and flux errors, furnishing insights into the strengths and potential constraints of our proposed method.

(i) Peak Signal-to-Noise Ratio (PSNR) signifies the peak error in reconstruction quality in image compression. It quantifies this error by computing the logarithm of the ratio between the highest pixel value (MAX) of the ground truth and the Mean Squared Error (MSE) observed between the ground truth and the test image, with the result expressed in decibels. In our experiments, the ground truth refered to images of individual galaxies before artificial blending for both data types, while the test images are the deblended images from the corresponding \emph{CAT-deblender}.
\begin{equation}
    \text{PSNR} = 20 \cdot \log_{10}\left({\text{MAX}}\right)-10 \cdot \log_{10}\left({\text{MSE}}\right) \, \text{dB}
    	\label{eq:quadratic_9}
\end{equation}

(ii) The Structural Similarity Index (SSIM) is widely recognized as a benchmark tool for measuring the similarity between two images, denoted as x and y. It operates by conducting essential statistical comparisons between these images. These comparisons focus on pixel-level attributes, including the means (\( \mu_x \) and \( \mu_y \)), standard deviations (\( \sigma_x \) and \( \sigma_y \)), and the covariance (\( \sigma_{xy} \)) of the images. Here, the mean values (\( \mu_x \) and \( \mu_y \)) reflect the average intensity of the image pixels, the standard deviations (\( \sigma_x \) and \( \sigma_y \)) convey the range of intensity variation, and the covariance (\( \sigma_{xy} \)) assesses the pixel-level relational dynamics between the images, thus indicating their structural resemblance. The formula incorporates two small constants, \( c_1 = (K_1L)^2 \) and \( c_2 = (K_2L)^2 \), designed to ensure computational stability especially when the denominator approaches zero. Typically, K1 and K2 are preset to 0.01 and 0.03 respectively, with L denoting the pixel values' dynamic range. This methodological framework allows SSIM to provide a detailed and quantifiable method for assessing the visual similarities between two images, where x and y refer specifically to the images being compared, not the geometric axes.
\begin{equation}
    \text{SSIM}(x, y) = \frac{(2\mu_x\mu_y + c_1)(2\sigma_{xy} + c_2)}{(\mu_x^2 + \mu_y^2 + c_1)(\sigma_x^2 + \sigma_y^2 + c_2)}
        	\label{eq:quadratic_10}
\end{equation}

(iii) The ellipticity is defined as \( |e| = \frac{(a - b)}{(a + b)} \) \citep{arcelin2021deblending}, where \( a \) and \( b \) are the lengths of the semi-major and semi-minor axes, respectively. The measurement of the semi-major and semi-minor axes is carried out on the convolved image of the \emph{DECaLS} three-band data using the function \texttt{sep.extract()}\footnote{\href{https://sep.readthedocs.io/en/v1.1.x/api/sep.extract.html\#sep.extract}{\texttt{sep.extract()}}} from the \texttt{SExtractor} package.

(iv) The magnitude is calculated based on the total flux, which is obtained by simply adding the number of photons for each pixel. The measurement is carried out on the convolved image of the \emph{DECaLS} band data using the function \texttt{sep.extract()} from the \text{SExtractor} package.
\subsection{Deblending performance in artificial blended galaxy dataset}
\begin{figure}
	\includegraphics[width=\columnwidth]{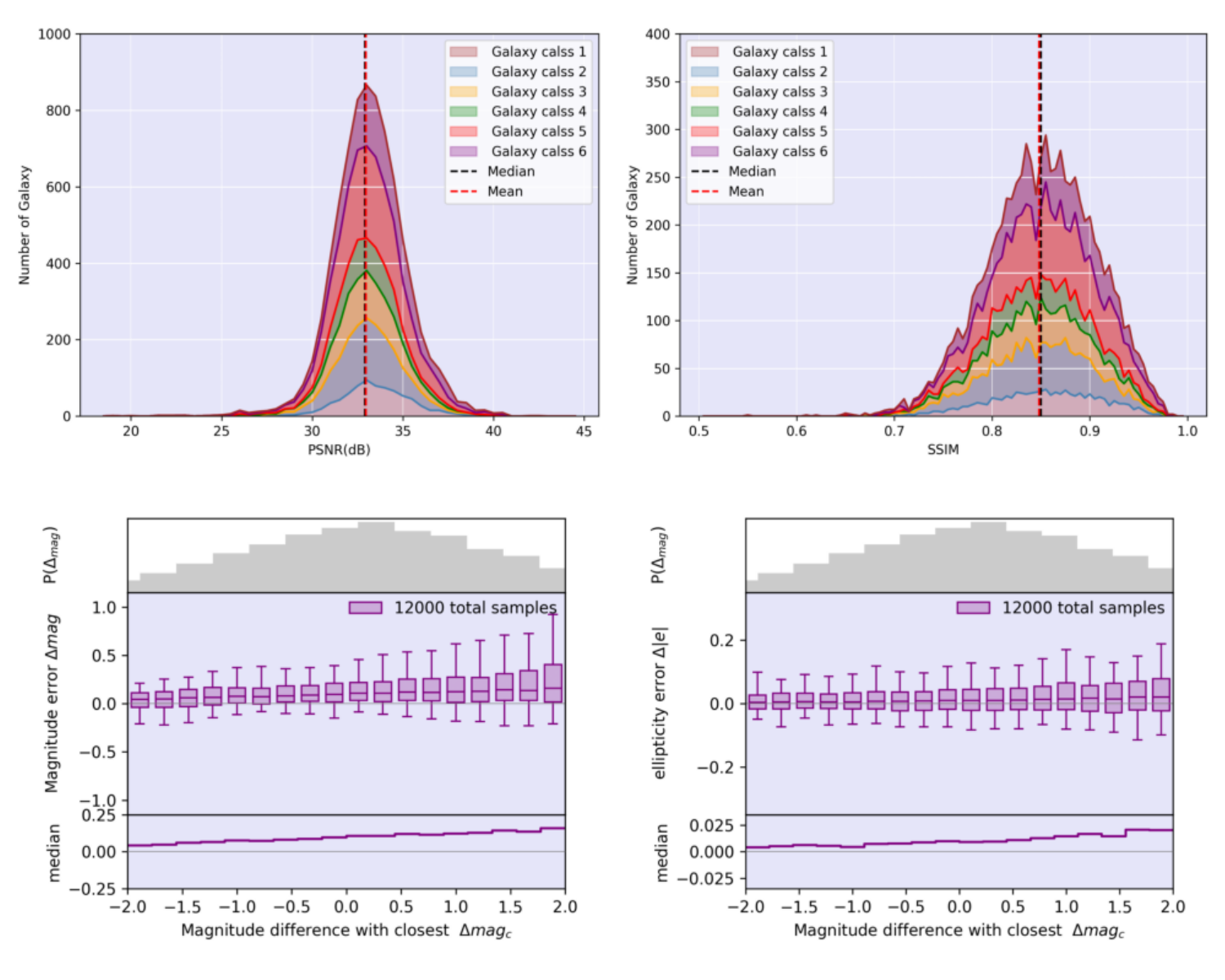}
    \caption{The two subfigures above show the distribution of PSNR and SSIM scores on the RGB data test set. The average PSNR score on the test set is 32.94 dB, with a standard deviation of about $\sigma = 4.06$ The maximum SSIM score on the test set is 0.982, with a standard deviation of about $\sigma = 0.003$. The lower two subfigures illustrate how brightness differences between the central and neighboring galaxies in the \emph{r}-band test images affect ellipticity and galaxy magnitude error recovery, the top panel shows the distribution of the number of test samples on the x-axis, the middle panel shows the distribution of errors when splitting the test samples in a specific number of bins, and the bottom panel shows the median error.}
    \label{fig:example_figure_5}
\end{figure}

For evaluating our model with RGB images, we utilize blended galaxy images in RGB from the test dataset for model inference. Since we maintained the original, unblended central galaxies in these composites, it has become feasible to conduct image comparisons using relevant metrics. Throughout the training program, the \emph{CAT-deblender} has neither encountered these particular combinations nor the possible individual galaxies. Within the test dataset, we computed the PSNR and SSIM indices for galaxies with varying morphologies. Additionally, we derived the mean, median, apex, nadir, and variance of their collective distribution. The average PSNR score was 32.94 dB ($\sigma$ = 4.06) and SSIM scores peaked at 0.982, averaging 0.849 ($\sigma$ = 0.003). The exhaustive distributions are illustrated in the two upper subfigures of Figure~\ref{fig:example_figure_5}.

For the multi-band images, we used the same neural network model to perform deblending as we did with the training of the RGB images, with the same architecture, and initial weights loaded from the previously trained prior model. To evaluate our method, we extracted mixed multi-band images of galaxies from the test set for inference. Similarly, for these overlapping galaxies, we could access the original pre-mixed isolated central galaxies, therefore, image comparison can be conducted with relevant metrics. We tested 12,000 mixed galaxies, including 4,000 dual-mix galaxies, 4,000 tri-mix galaxies, and 4,000 quad-mix galaxies. We undertook a comprehensive analysis of the error distribution in ellipticity and magnitude for target central galaxies within a multi-band imaging test dataset, both pre and post-deblending. It was observed that the peak signal-to-noise ratio's decrement markedly accentuates both the median error and its associated variance. Figure~\ref{fig:example_figure_5} elucidates, in its two lower subfigures, that in the r-band of the test dataset, the median error for ellipticity is consistently bounded within ±0.025, whereas the median error for the target galaxy magnitude does not exceed ±0.25. The impact of magnitude disparity between the target central galaxy and its nearest eccentric neighboring galaxy on the model's performance was analyzed. The lower two subfigures of Figure~\ref{fig:example_figure_5} also show that an increase in luminosity disparity of the nearest eccentric galaxy relative to the central galaxy leads to greater restoration errors in the magnitude and morphology of the central galaxy.  Specifically, a higher luminosity in the eccentric galaxy exerts a more substantial impact on the deblending model's performance.

\begin{figure*}
	\includegraphics[width=\textwidth]{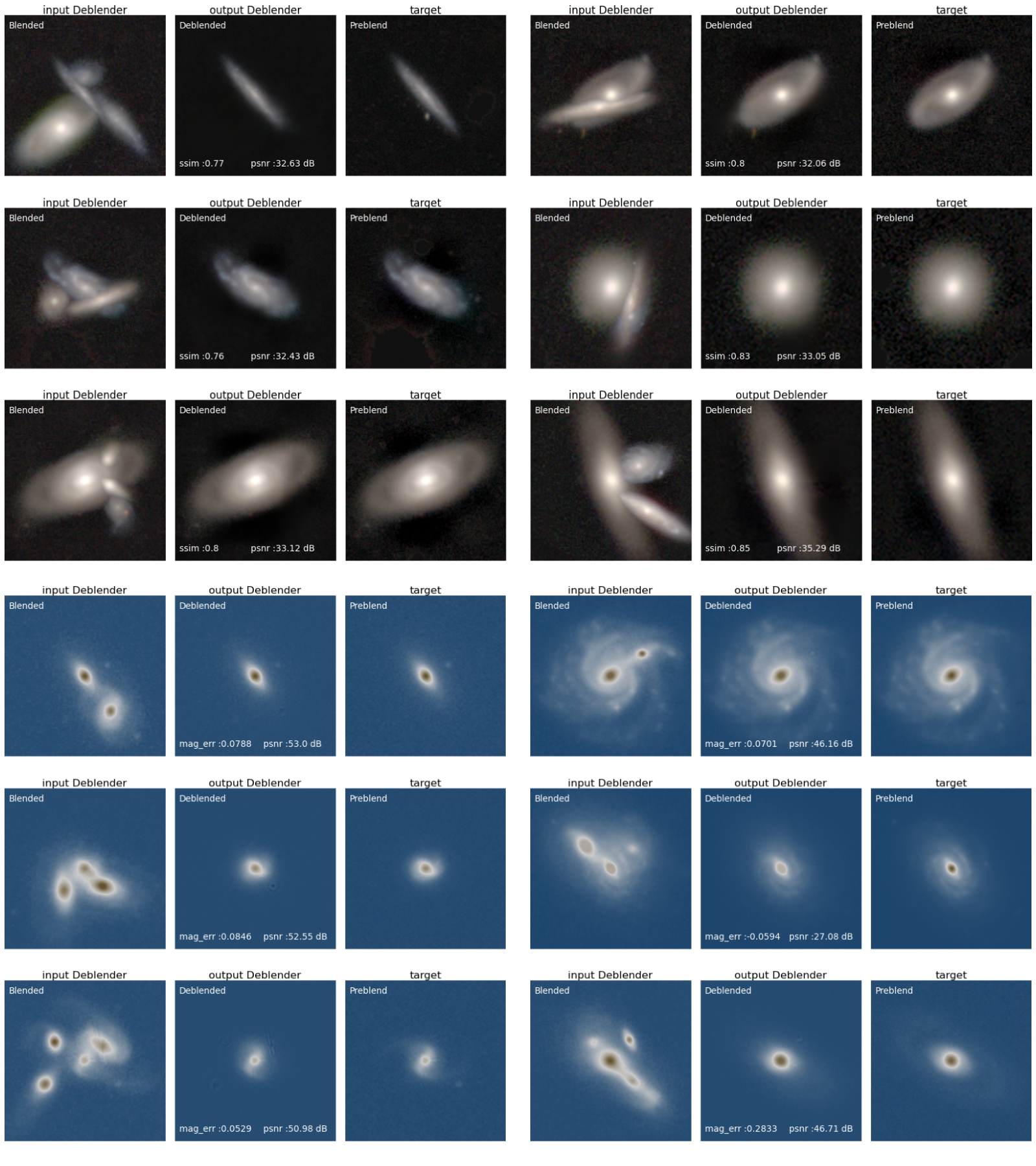}
    \caption{The examples of successful deblending and prediction on the test set for two data types are shown via \emph{CAT-deblender}. The top three rows display examples of successfully deblended RGB images, while the bottom three rows show examples of successfully deblended \emph{r}-band images in the multi-band data. In each panel, the left side represents the input blended galaxy image, the middle represents the deblended image predicted by the deblending model, and the right side represents the image of the pre-blended central target galaxy. Additionally, overlayed on the deblended predictions are the correlation PSNR score, magnitude measurement error, and SSIM score between the deblended image and the corresponding pre-blended image. Under various galaxy morphologies and any number of off-centered galaxy instances, our \emph{CAT-deblender} successfully deblended the central target galaxy and provided its most likely flux based on the off-centered galaxy occlusions.}
    \label{fig:example_figure_6}
\end{figure*}

Figure~\ref{fig:example_figure_6} displays examples of successful deblending for the two data types in the test set, along with their respective deblending performance metrics. Specifically, it includes the PSNR and SSIM scores for target galaxies in RGB images before and after deblending, as well as the measurement errors of target galaxy magnitudes and the PSNR values for target galaxies in multi-band images pre and post deblending.

\begin{figure*}[!t]
	\includegraphics[width=\textwidth]{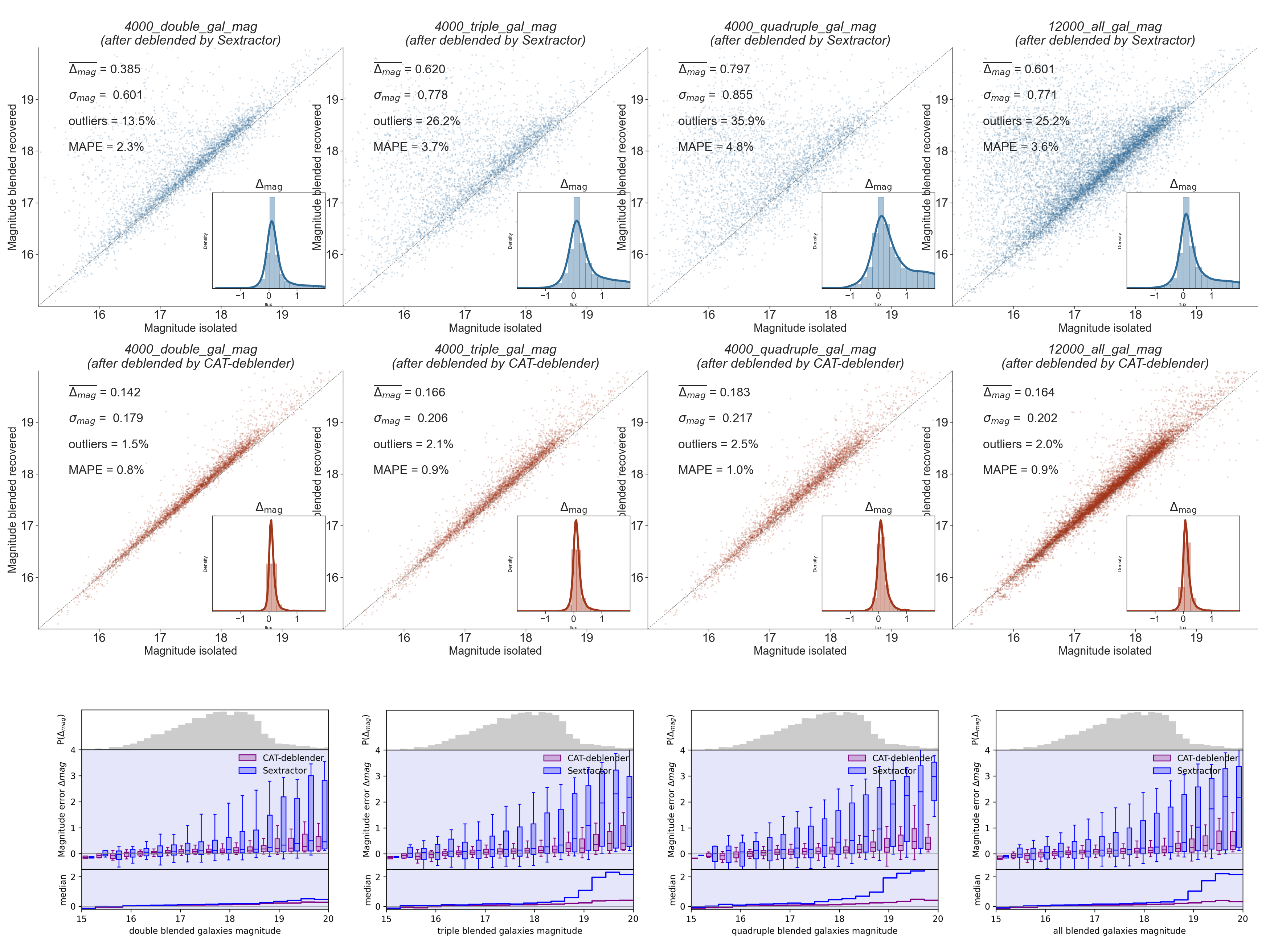}
    \caption{The first row of subfigures displays the magnitude error of central galaxies in blended galaxies measured using the direct deblending functionality of SExtractor, compared to their original magnitudes. The second row shows the magnitude error of central galaxies after deblending with CAT-deblender, in comparison to their original magnitudes. Each panel in the first two rows presents the average luminosity error (\(\Delta mag\)), dispersion (\(\sigma mag\)), outlier ratio (defined as \(|\Delta mag| > 0.75\)), and the Mean Absolute Percentage Error (\emph{MAPE}) of magnitude. The bottom row of four subfigures illustrates how the magnitude size of the central galaxy affects the recovery of galaxy magnitude in the model. In each subfigure, the top panel shows the distribution of the number of test samples along the x-axis, the middle panel shows the distribution of errors when dividing the test samples into a specific number of bins, and the bottom panel shows the median error. The first three subfigures in each row sequentially represent scenarios with 2-4 blended galaxy samples, and the fourth subfigure represents the entire test set sample.}
    \label{fig:example_figure_7}
\end{figure*}
We compared our deblending strategy with \emph{SExtractor} which is widely considered the industry standard and has been the baseline for detection, deblending, and image extraction in astronomy for more than 20 years. It returns a set of user-specified parameters from a broader default parameter file by following user-defined specific configurations. We noted that this is an inherently unfair comparison for two reasons. First, it has been trained on this specific set of galaxies, and thus incorporates priors adjusted correctly for the morphological, flux, and ellipticity distribution, while \emph{SExtractor} uses a general algorithm, and is therefore essentially more robust. The deblending method of \emph{CAT-deblender} is to strip single objects, while \emph{SExtractor} detects objects by creating a segmentation map. Despite this, it is a reasonable sanity check to measure what improvements can be brought about by using more complex methods. In Table~\ref{tab:example_table_2}, we present a comparison of three deblenders – SExtractor \citep{bertin1996sextractor}, GAN(Generative Adversarial Network) \citep{reiman2019deblending}, and CAT-deblender – focusing on their ability to restore the central galaxy's magnitude and ellipticity, as well as their performance in terms of Structural Similarity Index (SSIM) and Peak Signal-to-Noise Ratio (PSNR). Figure~\ref{fig:example_figure_7} offers a comparative analysis between \emph{CAT-deblender} and \emph{SExtractor} in the realm of deblending proficiency, specifically pertaining to the magnitude measurements of the targeted central galaxies. The panels in the first quadrant display the variation arising from SExtractor's deblending, comparing the central galaxy's measured magnitude post-deblending (y-axis) with its intrinsic magnitude before blending (x-axis). In contrast, the second quadrant panels underscore the differences in the central galaxy's magnitude after deblending with CAT-deblender (y-axis) against its original magnitude prior to blending (x-axis). The third row of Figure ~\ref{fig:example_figure_7} displays the residual magnitudes, which represent the magnitude recovery errors, for both SExtractor and CAT-deblender across the test set at various magnitude levels. We observe that as the magnitude increases, both deblenders exhibit a performance decline across various numbers of blended galaxies, showing a tendency to assign lower flux values to the identified central galaxies. However, compared to SExtractor, CAT-deblender does not exhibit a significant decrease in the accuracy of central galaxy magnitude recovery as the number of blended galaxies increases. In all cases, CAT-deblender outperforms SExtractor, with the median reconstruction error for r-band consistently lying within ±0.25. Even in the presence of three eccentric galaxies, CAT-deblender's Mean Absolute Percentage Error (MAPE) remains a modest 1\%, markedly lower compared to SExtractor's 4.8\%. 

\begin{figure}[!t]
	\includegraphics[width=\columnwidth]{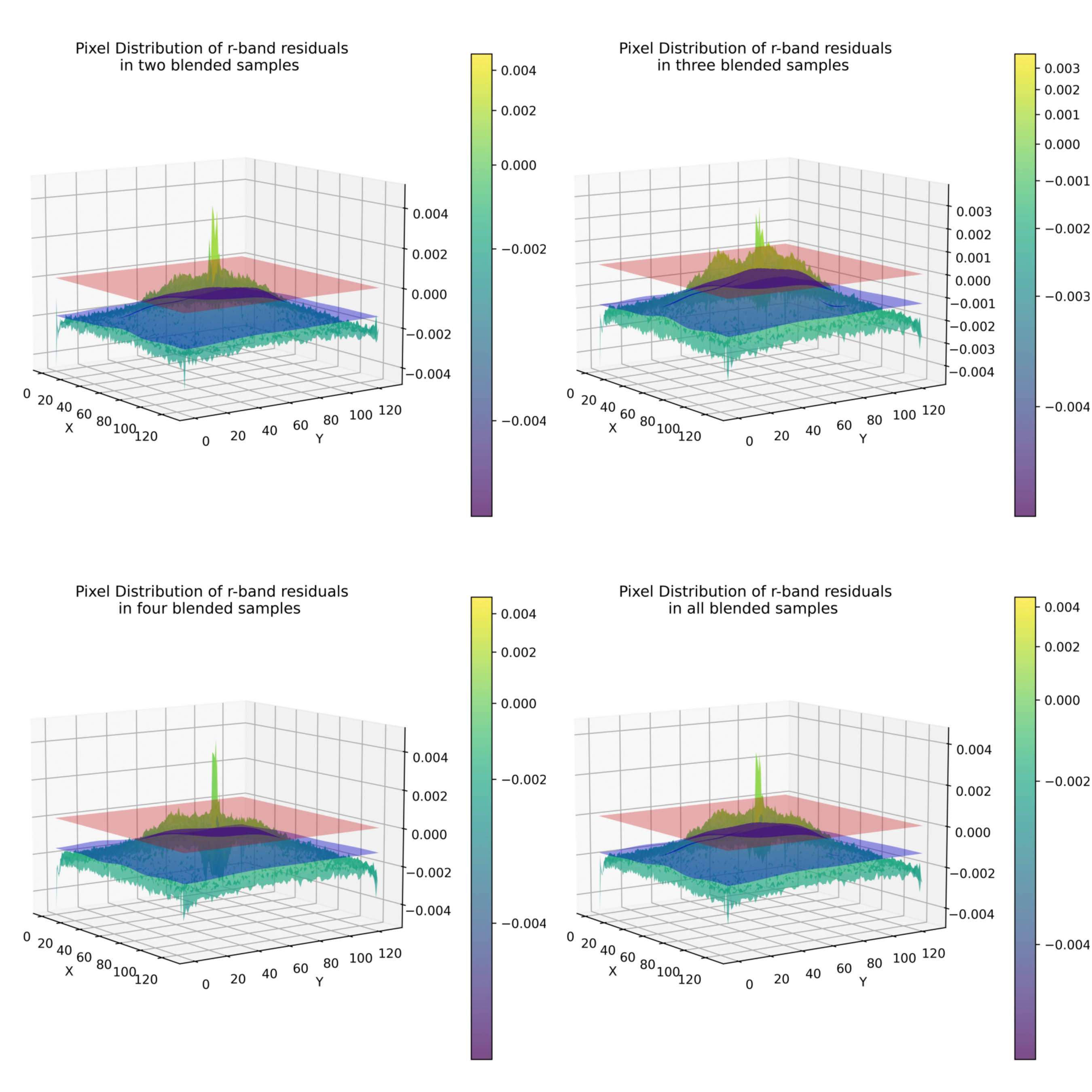}
    \caption{The average residual pixel distribution of the \emph{r}-band image unmixing on the test set is shown. The top two subfigures display the unmixing average residuals for the bi-blended, tri-blended galaxies, while the bottom two subfigures display those for the quad-blended galaxies and all samples in the test set. 
For each subplot, the x and y axes represent the height and width of the image, the bar on the right represents the residual pixel values. The red semi-transparent plane above represents the pixel zero level, and the blue semi-transparent surface below represents the background pixel distribution in the residual images.}
    \label{fig:example_figure_8}
\end{figure}

\begin{table}[!t]
\centering
\caption{The comparison of the three deblenders, SExtractor \citep{bertin1996sextractor}, GAN(Generative Adversarial Network) \citep{reiman2019deblending}, and CAT-deblender, in terms of restoring the central galaxy's magnitude, ellipticity, as well as their performance in SSIM and PSNR, is shown in the following table. In this table, Mag\_mape stands for the Mean Absolute Percentage Error between the magnitude of the deblended galaxy and the original central galaxy's magnitude. $\Delta mag$ represents the average magnitude error, and $\Delta ellipticity$ indicates the average error in ellipticity. Both are calculated as the difference between the predicted value and the actual value.}
	\label{tab:example_table_2}
\begin{tabular}{cccccccc}
\hline
Deblender & Mag\_mape     & $ \Delta mag $         & $ \Delta ellipticity $          & SSIM & PSNR\\
\hline
SExtractor & 3.61\% & 0.52 & 0.056 & -- & --\\
GAN & 1.79\% & 0.32 & 0.019 & 0.96 & 42.65\\
CAT-deblender & 0.9\% & 0.13 & 0.015 & 0.99 & 48.79\\ 
\hline
\end{tabular}
\end{table}
To provide a two-dimensional perspective on the model's performance, we evaluated the average residuals from \emph{r}-band deblending across a range of multi-band test images. Here, \text{residual} denotes the pixel difference between the target central galaxy image and its post-deblending version produced by our model. Figure~\ref{fig:example_figure_8} displays these residuals for the \emph{r}-band within multi-band images. This encompasses a sample set totaling 12,000 images—4,000 each of bi-blended, tri-blended, and quad-blended galaxies. Empirically, the \emph{r}-band image residuals consistently remained below 0.005, which is substantially less than the typical \emph{r}-band pixel threshold. Despite this overall performance, heightened residual pixel values were noticeable at the core regions of the target central galaxies. Furthermore, the predominant negative shift in the background of our residual images can be traced back to our synthetic images containing a double-layered background. This dual-background feature becomes apparent post-deblending due to the model's inclination to emphasize galaxies with higher pixel values. Another key observation is the model's consistent performance amidst increasing counts of eccentric galaxies. The residual pixel values for target central galaxies remain minimal even as the number of such galaxies rises. 
\begin{figure*}
	\includegraphics[width=\textwidth]{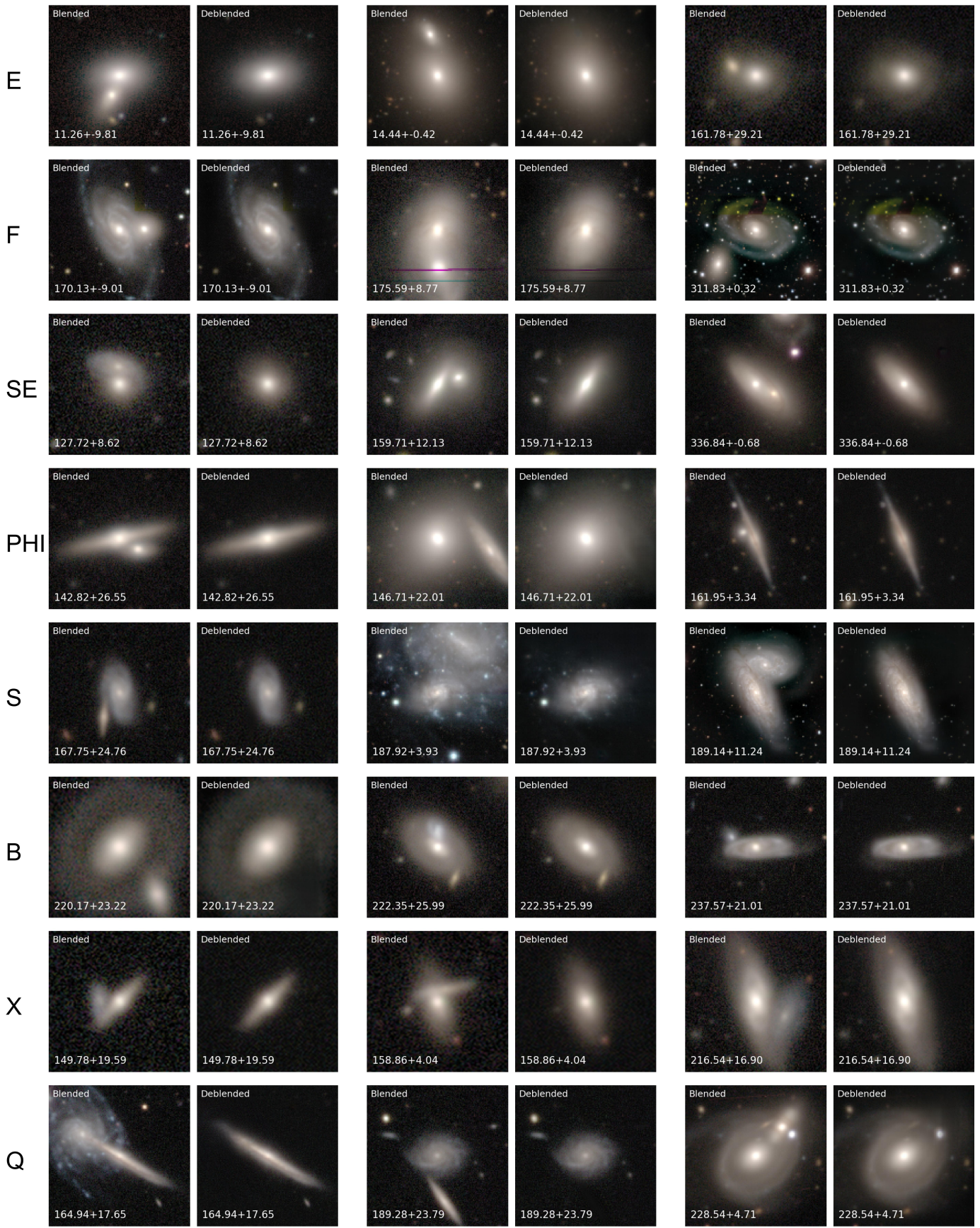}
    \caption{We present examples of RGB images of various overlapping galaxy types from the DECaLS database, along with the deblending results obtained using the \emph{CAT-deblender} model. Each subplot consists of a pair of images. On the left side, we show the real observed blended galaxy image, and on the right side, we display the successful deblending prediction by the \emph{CAT-deblender} model. Truncated coordinates are provided for identification purposes. The characters on the left side of each panel represent different blended galaxy types, and further details can be found in \citet{keel2013galaxy}.}
    \label{fig:example_figure_9}
\end{figure*}
\begin{figure*}
	\includegraphics[width=\textwidth]{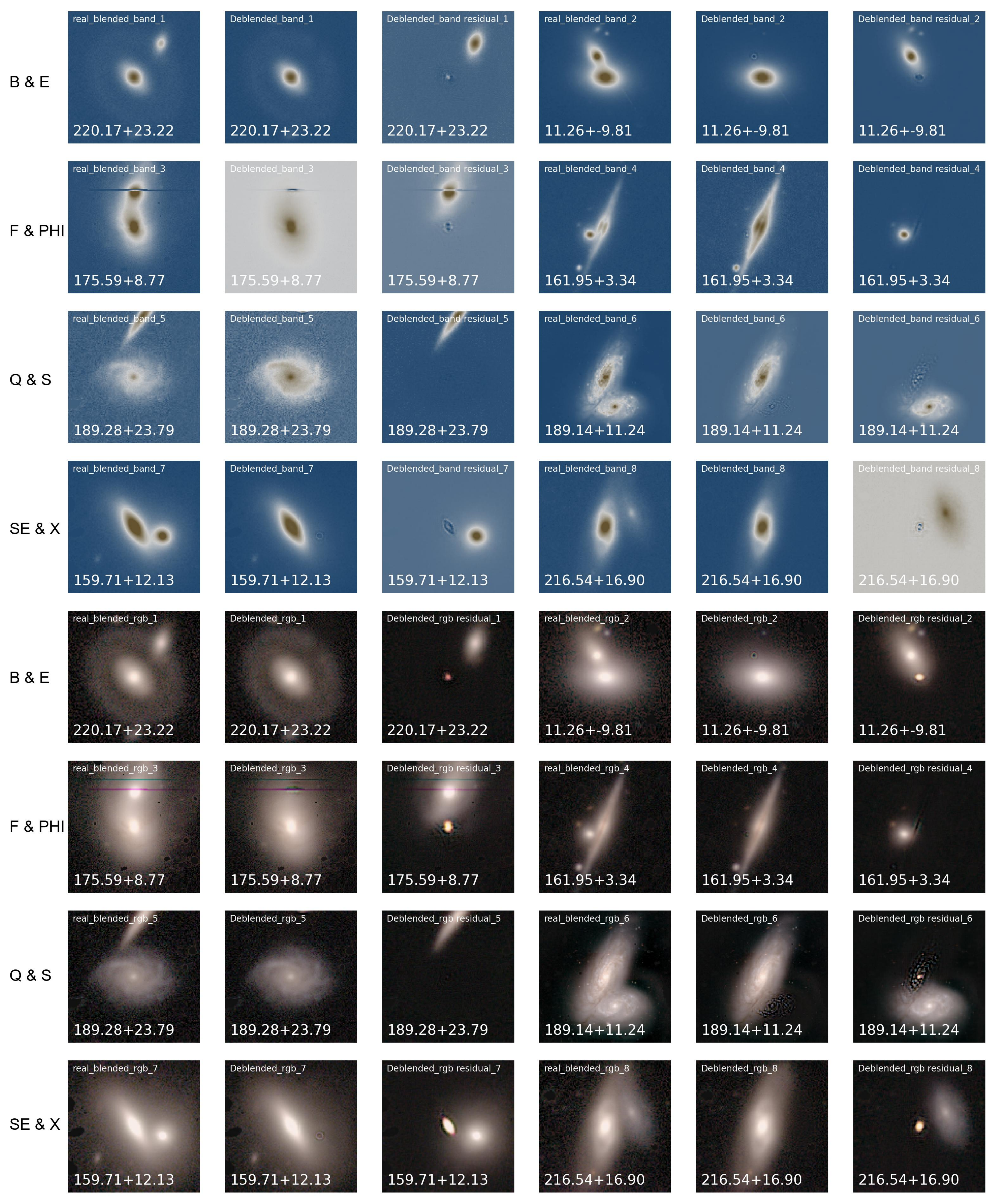}
    \caption{We present examples of different mixed galaxy types in the \emph{r}-band images from the DECaLS database, along with the deblending results obtained using the \emph{CAT-deblender} model. Examples of successful unmixing in the \emph{r}-band image from the multi-band images in the four panels at the top, and the four panels at the bottom represent successful unmixing examples in the corresponding RGB images for each of the four panels at the top. Each panel consists of three images. On the left side, we show the real observed mixed galaxy image. In the middle, we display the successful deblending prediction by the \emph{CAT-deblender} model. On the right side, we show the residual image obtained by subtracting the predicted galaxy pixels from the original mixed galaxy image. Truncated coordinates are provided for identification purposes. The characters on the left side of each panel represent different mixed galaxy types, and further details can be found in \citet{keel2013galaxy}.}
    \label{fig:example_figure_10}
\end{figure*}
\subsection{Deblending of blended galaxies in the DECaLS database}
In this section, we employed the pretrained \emph{CAT-deblender} model to attempt the deblending of galaxy images from \emph{DECaLS} database, encompassing both RGB and \emph{grz}-band. To ensure accuracy, we performed a cross-match between the \emph{SDSS} DR7 catalog of blended galaxies \citep{keel2013galaxy} and the GZD-5 morphological catalog, identifying a consistent set of 480 blended galaxies. From this intersecting dataset, the \emph{CAT-deblender} algorithm successfully deblended 433 galaxy pairs. To further evaluate our model's deblending capabilities on a larger scale, we utilized \emph{SExtractor} to search for blended galaxies in the DECaLS database. During this process, we deployed the \texttt{sep.extract()} function of the \emph{SExtractor} software for astronomical source detection and mask generation. We examined whether there are adjacent masks around the mask containing the central pixel of the image; if present, it indicates that an eccentric galaxy is blended with the central galaxy, and we categorize such pairs of galaxies as blended galaxies. This allowed us to filter blended galaxies within the \emph{DECaLS} database. After the deblending process with our model, we could use the deblending feature of \emph{SExtractor} to verify the deblending effectiveness. To validate the robust deblending capability of our model, we attempted to deblend 100,000 randomly selected blended galaxy images from the \emph{DECaLS} database. After verification using SExtractor's deblending, all the successfully deblended galaxies were subsequently cataloged. Herein, we present only a subset of our deblending results. A selection of successfully deblended entries can be viewed in Table~\ref{tab:example_table_3}, while a more extensive deblending catalog is hosted on our \href{https://github.com/SDU-ZR/CAT-deblending}{GitHub} repository, accessible via the link provided below.

First, we preprocessed the blended galaxy data in the same manner as the training set data, and then input it into the trained model for inference. Figures~\ref{fig:example_figure_9} and~\ref{fig:example_figure_10} display the results of successfully deblending real mixed galaxies. Figure~\ref{fig:example_figure_9}  shows the deblending results of 8 types of blended galaxies using the RGB image deblending model; the left side represents the blended galaxy images captured by actual survey projects, while the right side depicts the successfully deblended predicted images by \emph{CAT-deblender}. In several successful cases of deblending RGB blended galaxy images, our deblender, successfully segmented and deblended the central galaxies. It not only inputted the most likely flux for the pixels of the central galaxies obscured by eccentric galaxies, but also further restored the most probable morphology of the central galaxies. Figure~\ref{fig:example_figure_10} presents the deblending effects of the \emph{r}-band images after 8 different blended galaxies have been processed by the multi-band image deblending model, as well as their conversion to RGB images. From Figure~\ref{fig:example_figure_10}, we observe that CAT-deblender also demonstrated excellent performance in deblending multi-band images, effectively restoring the central galaxies to their most probable flux values and original morphologies in each band, and additionally showing effective color restoration in the converted RGB images. In this figure, we also display the residual images after deblending. We notice that the residuals at the core positions of the original central galaxies in the residual images are the most pronounced, which is consistent with the residual analysis results from the test set of artificially blended galaxies.

\begin{figure*}
	\includegraphics[width=\textwidth]{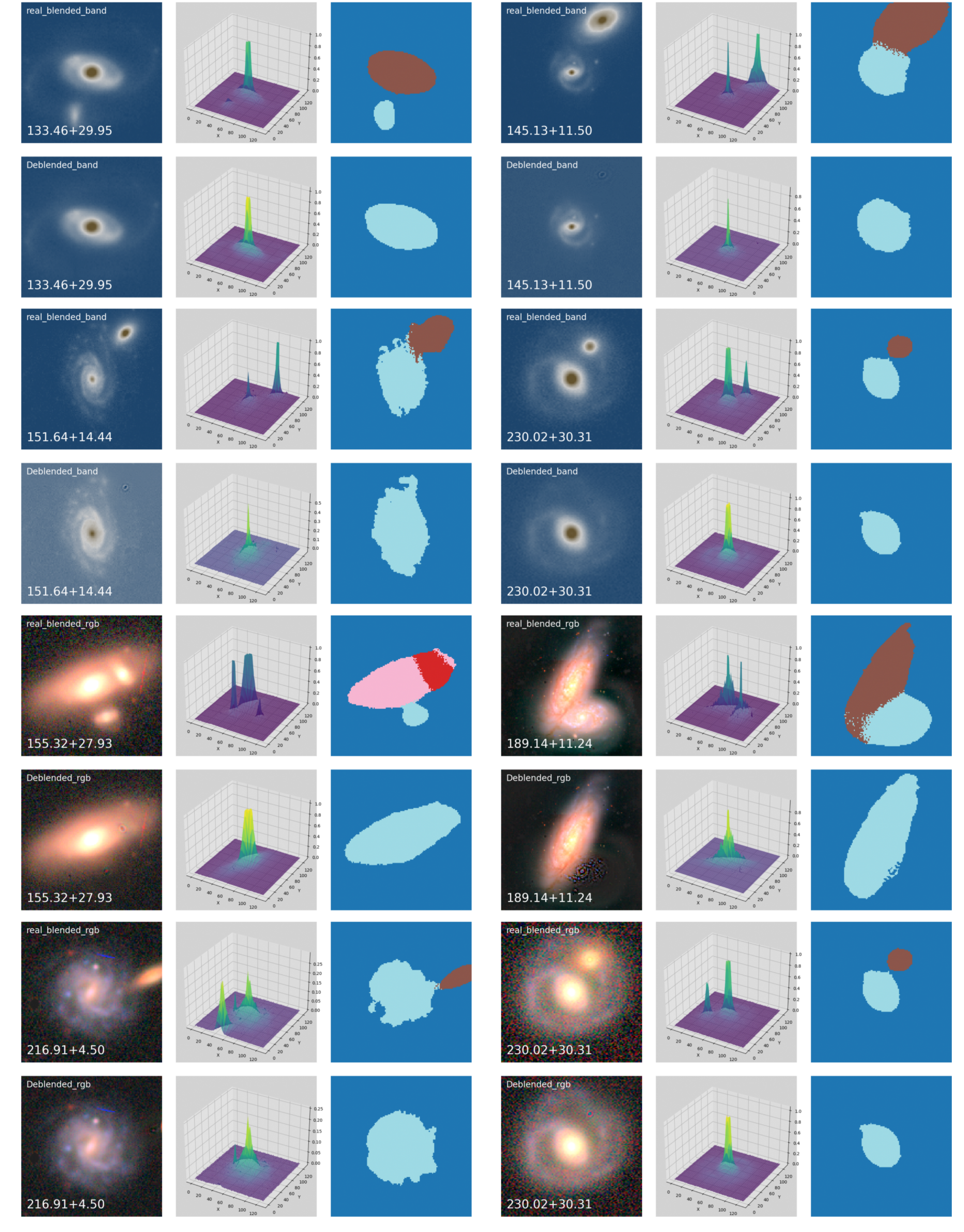}
    \caption{We present a subset of successfully deblended data types selected from the DECaLS database. For each subplot, the top three panels represent the mixed galaxy image, the three-dimensional pixel value distribution of the mixed galaxy image, and the deblending mask obtained using \emph{SExtractor}. The bottom three panes represent the deblended target galaxy image, the three-dimensional pixel value distribution of the deblended image, and the deblending mask obtained using \emph{SExtractor}.}
    \label{fig:example_figure_11}
\end{figure*}

To verify the success of galaxy deblending, we combined the use of \emph{SExtractor} and manual visual inspection. As illustrated in Figure~\ref{fig:example_figure_11}, the top three panels of each subplot respectively display the blended galaxy image, the three-dimensional pixel distribution of the blended galaxy image, and the deblending mask from \emph{SExtractor}. The bottom three panels represent the deblended isolated target galaxy image, the three-dimensional pixel distribution of the image after deblending, and SExtractor's deblending mask. By comparing the before and after images and analyzing the three-dimensional pixel distribution graphs, it is evident that the pixel flux and original morphology of the central isolated galaxy have been restored after deblending. This validation of the deblending functionality also confirms the successful deblending capability of our model.

\begin{figure*}[!t]
	\includegraphics[width=\textwidth]{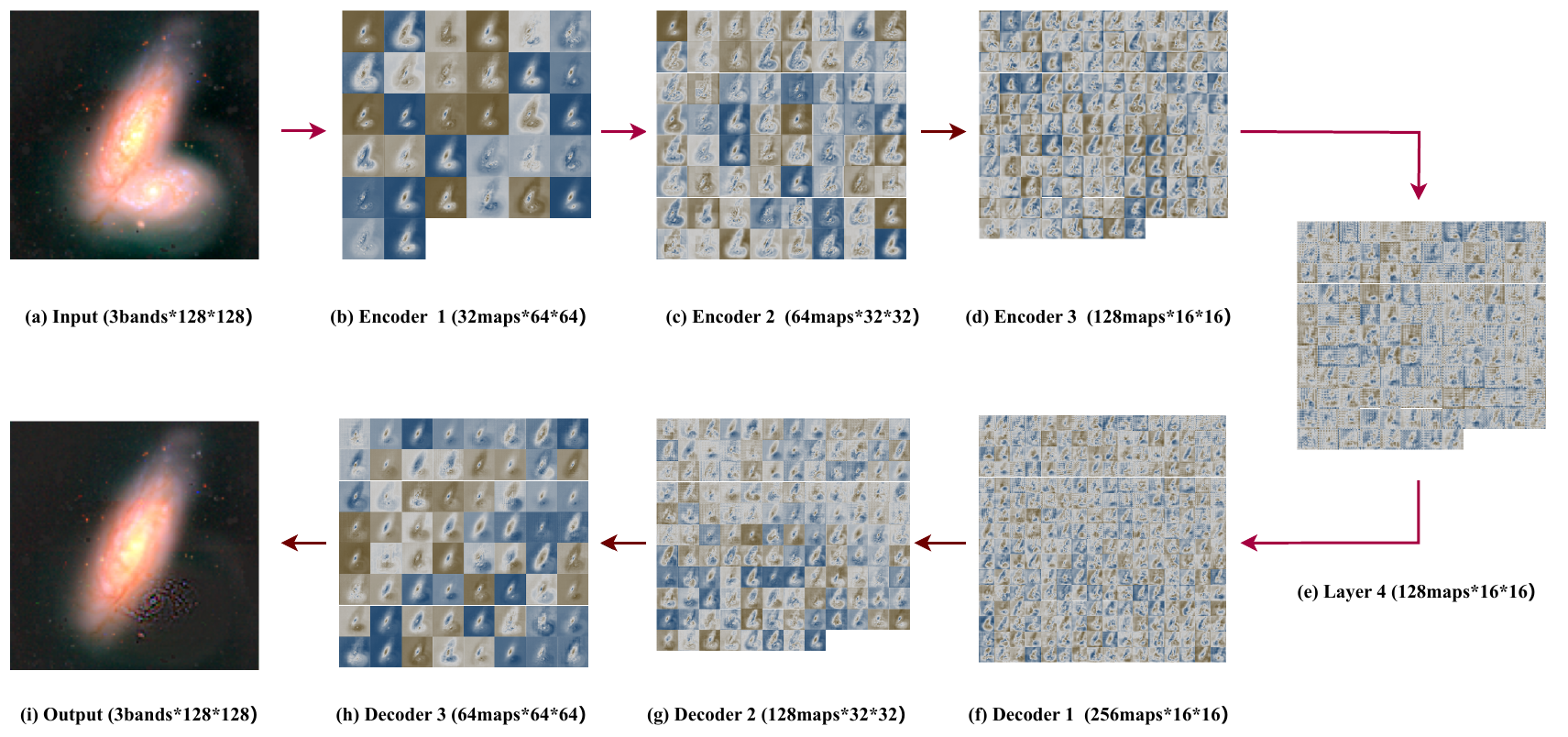}
    \caption{(a) A 128×128 three-band blended galaxy image sample from the DECaLS dataset, visualized through Lupton transformation \citep{lupton2004preparing}. (b): Activation on the first layer of the Encoder when a 3×128×128 image is input into the network. (c) Activation on the second layer of the Encoder. (d) Activation on the third layer of the Encoder. (e) Activation on the fourth layer, the bottleneck layer. (f): Activation on the first layer of the Decoder. (g) Activation on the second layer of the Decoder. (h) Activation on the third layer of the Decoder. (i) A 128×128 three-band deblended galaxy image sample, visualized through Lupton transformation \citep{lupton2004preparing}. Images in (b) to (h) represent a feature map each; the Encoder and bottleneck layers correspond to the output of learned features, while the Decoder layers correspond to the output of reconstructed features.}
    \label{fig:example_figure_12}
\end{figure*}
\begin{table*}[!t]
\centering
\caption{The model successfully deblending a catalog of blended galaxies is shown.}
	\label{tab:example_table_3}
\begin{tabular}{cccccccc}
\hline
SDSS\_ID & DECaLS\_iauname     & Ra          & Dec          & Type & Redshift    & R\_mag    & Sersic\_nmgy\_r \\
\hline
587722981748637919 & J131820.01-010530.0 & 199.583403  & -1.091679248 & S    & 0.085784696 & 16.101273 & 697.34125   \\    587722983901954254 & J141132.36+002756.8 & 212.8848652 & 0.465784781  & E    & 0.14471209  & 16.707241 & 422.78766       \\
587726033851252911 & J104747.14+032011.1 & 161.9463595 & 3.336409926  & Phi   & 0.04288498  & 16.379517 & 470.9149 \\
587726031722184751 & J133558.58+014348.2 & 203.994091  & 1.730047291  & Q     & 0.022911945 & 14.392784 & 2556.1475\\
587734303805604056 & J222721.06-004041.1 & 336.837751  & -0.678105121 & SE    & 0.05659264  & 15.034672 & 1929.8839\\
587736584973189140 &J154424.00+293416.4  &236.1000451  & 29.57123935  & F     & 0.059020456 & 17.428226 & 199.35558\\
587724649255141394 & J114224.33-023406.7 & 175.6014229 & -2.568541911 & X     & 0.043405034	& 17.471579	& 118.15414\\
588848899916562639 &J123538.68-001221.6  & 188.9112479 & -0.205992728 & B	  & 0.022795824	& 15.429775	& 765.5961 \\
 ...               & ...                 & ...         & ...          & ...  & ...         & ...       & ...     \\
\hline
\end{tabular}
\end{table*}
Visual analysis of neuron activations in the transformer provides insights into its processing dynamics. Figure~\ref{fig:example_figure_12} depicts the neuron activation across each transformer layer when a blended galaxy image is introduced into the network. In the Encoder phase, the primary task is feature extraction. As we move deeper into the layers, the feature map dimension reduces. Initial layers possess fewer filters, gradually increasing in subsequent layers. The early feature maps predominantly capture rudimentary attributes like edges, colors, and basic textures. However, as we progress in depth, these maps evolve, encoding more sophisticated features such as distinct object parts, unique shapes, or intricate textures. The intermediary feature layer, bridging the Encoder and Decoder, encapsulates abstract representations. These might delineate nuanced details about the galaxy's structure or classification but are typically not intuitively discernible to humans. Transitioning to the Decoder phase, it commences with these abstract features, progressively refining and restoring details. Its intermediate layers amalgamate data from the Encoder, embarking on the rejuvenation of specific galaxy structures and nuances. By the culmination of the Decoder phase, the feature maps should bear a strong visual alignment with the deblended image, encapsulating crucial visual components.

\begin{figure*}[!t]
	\includegraphics[width=\textwidth]{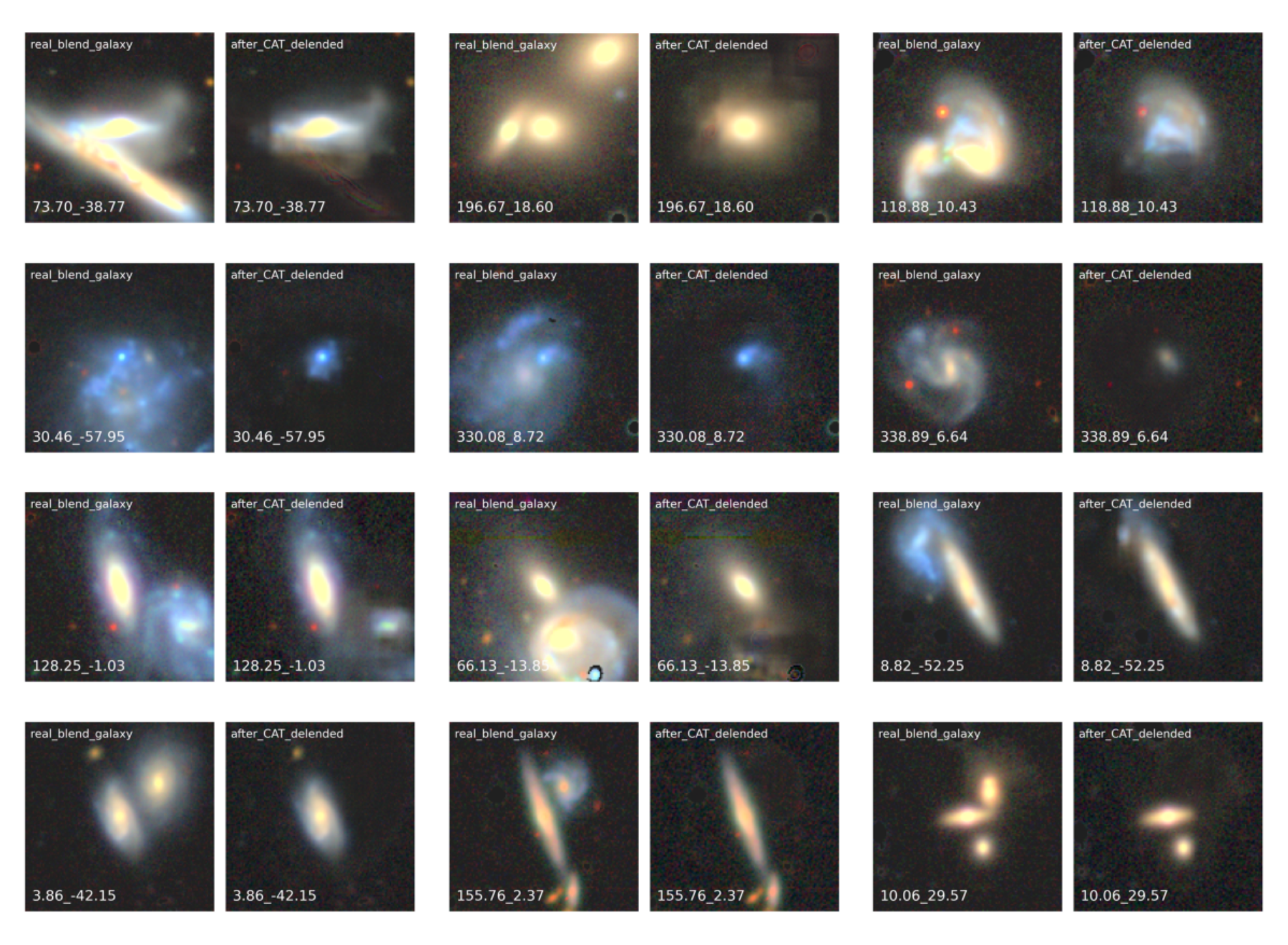}
    \caption{For those images that were successfully deblended but still have issues, we observed two distinct patterns: (i). There is a missing overlap area between the central galaxy and the eccentric galaxy in the deblended image, where the pixels are not appropriately filled, as shown in the three subfigures of the first row. (ii). The peak position of the central galaxy deviates significantly from the center of the image, resulting in a false deblend, as illustrated in the three subfigures of the second row. For those images that were not successfully deblended, we also found two common patterns: (i): The target image still retains noticeable remnants of the eccentric galaxy, as shown in the three subfigures of the third row. (ii): While there are multiple eccentric galaxies surrounding the central galaxy, the deblended image still retains parts of the eccentric galaxies, as depicted in the three subfigures of the fourth row.
}
    \label{fig:example_figure_13}
\end{figure*}

From the \emph{DECaLS} database, we identified 100,000 blended galaxy samples using the \emph{SExtractor} tool. Upon rigorous preprocessing, these samples were subjected to CAT-deblender's multi-band image deblending model. Notably, post-inference, we verified effective deblending in 63,733 galaxy images. In this context, \emph{effective deblending} denotes instances where the post-deblended images allowed \emph{SExtractor} to recognize only a solitary galaxy. For those images that were successfully deblended but still had problems, we observed two different patterns: (i) The overlapping regions of the center and eccentric galaxies in the deblended image appeared to be missing and were not filled with the proper pixel values, as shown in the three subfigures of row 1 of Figure~\ref{fig:example_figure_13}. (ii) The peak position of the center galaxy is significantly off the center of the image causing pseudo deblending, as shown in the three subfigures of row 2 of Figure~\ref{fig:example_figure_13}. For those images that did not complete the deblending, we similarly found two general patterns: (i) The target image is left with a clear residual shadow of the off-center galaxy, as shown in the three subfigures of row 3 of Figure~\ref{fig:example_figure_13}. (ii) There are multiple eccentric galaxies around the central galaxy, while some of the eccentric galaxies remain in the deblended image, as shown in the three subfigures of row 4 of Figure~\ref{fig:example_figure_13}.
\section{Summary and Discussion}\label{sec:summary and discussion}
In this research, we introduce a novel methodology using the Transformer as the key neural network module. After deblending, our approach effectively recovers the flux and ellipticity of the target galaxy. Unlike previous studies, our method uses both RGB and \emph{grz}-band images from the \emph{DECaLS} database for training. A literature review shows that traditional deblending models mainly use simulated datasets for training. However, the performance of models trained only on simulated datasets for galaxy deblending in the \emph{DECaLS} database is uncertain. We use galaxy images from the \emph{DECaLS} database, free of unrelated astronomical objects, as the foundation for our network model training. Following the principles of galaxy target detection, we use identified galaxy peak centers to adjust image dimensions, addressing the galaxy deblending challenge head-on.

We set the input dimensions of both data types to $128\times128$ pixels, showcasing the efficacy of our \emph{CAT-deblender} model on the test set. After deblending, the RGB images had an improved signal-to-noise peak and consistently higher structural similarity to the ground truth. The average PSNR score was 32.94 dB ($\sigma$ = 4.06) and SSIM scores peaked at 0.982, averaging 0.849 ($\sigma$ = 0.003). In multi-band images, the ellipticity of central galaxies and the median reconstruction error for \emph{r}-band stayed within ±0.025 to ±0.25, indicating minimal pixel residuals. Our experiments revealed that brighter eccentric galaxies affected the \emph{CAT-deblender} model's performance more significantly. Therefore, in multiple galaxy blending, it's advantageous to prioritize deblending the brighter galaxy first. Compared to \emph{SExtractor}, our model maintained its performance even with increasing galaxy counts, indicating its suitability for denser galaxy fields. Even with three off-center galaxies blending with the central one, our model achieved a MAPE of just 1\% for the central galaxy's magnitude measurement post-deblending, emphasizing its potential for deblending in densely populated fields.

Using the \emph{CAT-deblender} model, we deblended 433 galaxies from the overlapping regions of \emph{DECaLS} and \emph{SDSS} catalogs, covering all combinations suggested by \citet{keel2013galaxy}. Using \emph{SExtractor}, we identified 100,000 overlapping galaxies in the DECaLS database. After validation with \emph{SExtractor} and manual inspection, our model effectively deblended 63,733 of these galaxies. These results highlight our model's capability for deblending overlapping galaxies in the DECaLS database. Trained on galaxy luminosity distributions, our model restores pixel values lost after blending. Our approach needs just one galaxy to be near the image center and efficiently processes large datasets from surveys like \emph{DECaLS} and \emph{SDSS}, enabling rapid deblending. 

 \emph{CAT-deblender} has not fully deblended the entire catalog of overlapping galaxies and tends to remove the most significant blended galaxies from the center, while leaving all smaller, more separated galaxies around it. On one hand, this may be attributed to our training dataset being based on data encompassing 2-4 blended galaxies orbiting a central galaxy. The inability to simulate all blending scenarios could limit its performance in handling blends with a greater number or more complex configurations of galaxies. On the other hand, the proportion of smaller, more dispersed galaxies in the training set is relatively low. Consequently, for those smaller or more dispersed galaxies which are less commonly featured in the training data, the deblender may struggle to accurately identify and process them. Additionally, we have identified a consistent positive bias in both the magnitude and ellipticity measurements, indicating that deblended galaxies exhibit an ellipticity and r-band magnitude higher than their actual values. This observation has significant implications for weak lensing measurements in two key aspects. Firstly, such a bias could lead to alterations in the weak lensing calibration factor, commonly referred to as the multiplicative bias. This factor is critically sensitive, with stringent accuracy requirements near 0.001 for projects like the Vera C. Rubin Observatory’s Legacy Survey of Space and Time (LSST). Secondly, the presence of a magnitude-dependent bias in the r-band magnitude might introduce distortions in photometric redshift (photo-z) estimations, particularly if there is a disparity in magnitude distributions between the spectroscopic redshift (spec-z) as the training set and photo-z datasets. These are crucial concerns that must be addressed in future developments if we aim to apply these findings effectively in the field of cosmology.
 
There are several areas our research identifies for improvement. Presently, we evaluate deblending effectiveness using tools like \emph{SExtractor} and manual reviews. Future research should concentrate on stringent assessment methods for authentic overlapping galaxies. The training model for our study primarily depends on the \emph{GZD-5} catalog, which includes predominantly lower redshift galaxies. However, galaxy morphology is intricately linked to redshift. Despite the potential morphological variations induced by galaxy redshifts, we observed a consistent pixel intensity pattern, wherein any redshifted galaxy is depicted as bright entities against a dark backdrop. Furthermore, our training set has a limited representation of smaller or more dispersed galaxies, with blending restricted to four or fewer, potentially leading to the deblender removing only the most significant blended galaxy from the central galaxies, while leaving all smaller, more dispersed galaxies intact. These findings highlight the potential of transfer learning in this domain, which we aim to further investigate in future research. Through ongoing iterative improvements, we intend to incorporate more advanced algorithms into practical astronomical research on galaxy deblending, thereby fostering continuous advancement in the field.

\begin{acknowledgement}
This research was supported by the Joint Research Fund in Astronomy under the National Natural Science Foundation of China (Grant No. U1931209), the Shandong Province Natural Science Foundation (Grant Nos. ZR2022MA089 and ZR2022MA076), and the Shandong Province Innovation Development Fund for Small and Medium-sized Science and Technology Enterprises (Grant No. 2022TSGC2492). We thank Professor Li Nan from the National Astronomical Observatories, Chinese Academy of Sciences, for their guidance. Our appreciation also extends to the reviewers for their insightful comments.
\end{acknowledgement}








\printbibliography

\appendix



\end{document}